
\hsize 15.0truecm \hoffset 1.0 truecm
\vsize 22.0truecm
\nopagenumbers
\headline={\ifnum \pageno=1 \hfil \else\hss\tenrm\folio\hss\fi}
\pageno=1

\def\lsim{\mathrel{\rlap{\lower4pt\hbox{\hskip1pt$\sim$}}
    \raise1pt\hbox{$<$}}}         
\def\gsim{\mathrel{\rlap{\lower4pt\hbox{\hskip1pt$\sim$}}
    \raise1pt\hbox{$>$}}}         

\hfill IUHET 248

\hfill DFTT 14/93

\line{\hfil April, 1993}

\bigskip
\vskip 12pt
\centerline{\bf Phenomenology of spin zero mesons and
glueballs}

\vskip 36pt
\centerline{Marco Genovese and Enrico Predazzi}

\centerline{\it Dipartimento di Fisica Teorica,
Universit\`a di Torino}

\centerline{\it and INFN, Sezione di Torino, I-10125,
Torino, Italy}
\medskip
\centerline{D. B. Lichtenberg}

\centerline{\it Department of Physics,
Indiana University, Bloomington, Indiana, 47405}

\vskip 1.0in
\item{} We discuss the phenomenology of
scalar and pseudoscalar mesons, emphasizing those which do
not carry manifest flavor quantum numbers. Many of
the properties of these mesons
are still not fully understood. Some of them
probably do not have the usual
two-quark (quark-antiquark) structure,
but may be mixed with glueball states or other exotics.
We construct or discuss simple models
for these mesons and point out which measurements
can shed light on their composition.

\vfill \eject
\centerline{I. INTRODUCTION}
\bigskip
Of all the low-mass mesons (masses below
2 GeV), the properties of the ones with spin zero
(both scalar and pseudoscalar)
are the least understood as
a class, especially those which do not carry
manifest flavor quantum numbers. We call
these latter
mesons ``flavorless,'' although some of them may have
hidden flavor. They are isoscalar mesons
which are self-conjugate (under charge conjugation).
Some of these mesons may be usual
two-quark (quark-antiquark) states,
but others may be four-quark states (two quarks, two
antiquarks), glueballs (composed only of gluons), or hybrids
(composed of a quark-antiquark pair plus glue). Of the
mesons with manifest flavor, some may be two-quark states,
but others may be four-quark or hybrid states.
Probably many of the observed spin-0
mesons are actually mixed states, and that is
what makes their structure so hard to determine.
We call attention to
three recent reviews of meson spectroscopy [1],
which discuss problems in distinguishing between
those mesons which are ordinary two-quark states and
those which are not.
Many of the known experimental properties of the spin-0
mesons are summarized in the tables of the
Particle Data Group [2].

Most of the light vector and tensor mesons
(containing only $u$, $d$, or $s$ quarks)
are ``ideally'' mixed,
which means that $SU(3)$ is broken in such a way that
the two physical flavorless states with isospin zero
have the quark composition $u\bar u + d\bar d$
and $s\bar s$.  A possible reason why light
flavorless spin-0 mesons are not ideally mixed
is that instantons contribute to  mixing them with glueball
states, as has been discussed by a number of
authors, including Shuryak [3],
Blask {\it et al.}\ [4], and Zakharov {\it et al.}\ [5].
The interaction induced by instantons is
apparently short range on the
scale of hadron size [6], and so should
be more important in the
pseudoscalar sector (in which the quark-antiquark pair
is in an $S$ wave) than in the scalar sector (in which the
quark-antiquark pair is in a $P$ wave).
Another possible reason
for non-ideal mixing of spin-0 mesons, whether carrying
manifest flavor or not, is mixing with four-quark and
and hybrid states.

In this paper we discuss the phenomenology of both
scalar and pseudoscalar mesons, emphasizing those
mesons which are flavorless and can mix with glueballs.
Some of the issues discussed are not new, but
they are presented in such a way as to allow a direct
comparison with models
once new data become available. We also try
to make clear where new measurements would
be particularly welcome and necessary for
testing theoretical schemes. Thus, throughout
the paper the emphasis is on the phenomenological
application of a large number of theoretical
ideas which in the previous literature
appear rather sparsely.
As a consequence, our paper contains not only new
results but some
old results presented in a rather new form.

Concerning the pseudoscalar states, our main approach
is not qualitatively different from
some previous ones [7],
although many details are different.
In some instances we simplify previous models;
this simplification enables us to make
a large number of predictions that can be tested
in the new generation of experiments designed to
search for glueballs.

In the scalar sector, we make a large number
of new  experimentally testable predictions plus
give an extensive resum\'e of old results
which, to the best of our knowledge, have never been
collected together previously. Among other things, our
ideas should allow future experiments to clear up most of
the puzzles concerning the low-lying $0^{++}$ resonances
and to disentangle the various associated ambiguities.
We also point out the limitations
of present theoretical schemes and cast our results in such
a form as to make clear where experiments
in progress will in the near future
be able to shed light on existing uncertainties.

Altogether, the principal new features, aside from
details,  are:
(i) We treat essentially all the light flavorless spin-0
mesons.
(ii) We emphasize models which have enough simplifying
features to enable us to make testable predictions.
(iii) We
point out
where the present theory is inadequate
and what experimental information is necessary to clarify
the situation.

The issues on pseudoscalar and on scalar glueball
candidates can sometimes
be sharply distinguished from each other.
Nevertheless, we have chosen to present
our results on scalars and pseudoscalars in a
single paper so as to present a view on spin-zero
glueball searches which is as unified as possible.

\bigskip
\centerline{II. THE PSEUDOSCALAR SECTOR}
\bigskip
\centerline{\bf A. Overview}
\medskip

In this section we discuss
flavorless pseudoscalar
mesons with positive $C$ parity.
We point out where our understanding is good,
where it falls short, and where experiments can improve
the situation. We need especially to improve our
understanding of glueballs. Although,
according to QCD, glueballs should exist,
theoretical calculations [8--22] of pseudoscalar glueball
masses vary over a large mass
interval, as can be seen from Table I.

The number of gluons in a glueball is not
necessarily
a conserved quantity in flux tube and string models and in
lattice QCD calculations, but in some models
the low-mass glueballs are composed either of two
gluons (digluonium) or three gluons (trigluonium).
The results of Table I are divided into flux tube and
lattice predictions and model predictions for
digluonium and trigluonium states. In this paper
we do not need to specify whether glueballs are
digluonium or trigluonium states. However, it is
plausible that the glueballs we discuss are digluonium
states, because the lowest digluonium is usually calculated
to be less massive than the lowest trigluonium state.
For a further discussion of trigluonium, see Anselmino
{\it et al.}\ [23] and references therein.

\medskip
\centerline{\bf B.  Simplified $\eta$, $\eta'$, $\iota$
mixing}
\medskip
Physicists have achieved remarkable
success in using quark potential models motivated by QCD
to predict the masses and other properties of light,
as well as heavy, mesons
as quark-antiquark bound states. See, e.g., the
paper of Godfrey and Isgur [24] and references therein.
However, without additional {\it ad hoc}
parameters, such models fail to give anywhere near
the correct masses of the $\eta$ and $\eta'$ mesons.

The problem of the $\eta$ and $\eta'$ mesons has been
with us a long time, and may arise in large part because of
instanton effects and also in part because of conventional
quark-antiquark annihilation diagrams. Both
these contributions are likely to lead to
$\eta$ and $\eta'$ mesons which contain not only
quark-antiquark pairs but also a gluonic component [25].
However, some experiments
by the DM2 [26] and MARK III [27] collaborations
indicate that the glueball
content of the $\eta$ is either absent or small.
A good candidate for a  state which is largely
glue is the $\eta(1440)$, but this state is probably
mixed with two-quark states.

A scheme that has been suggested to discuss these states
follows an idea which was first advocated by
Pinsky [28] who
argued that doubly disconnected diagrams may be important
in the discussion of $J/\psi$ decays. As a development of
this idea,
the MARK III [29] and
DM2 collaborations
[26] suggested
a scheme with disconnected diagrams
in which neither the $\eta$ nor the $\eta'$ has
any gluonic component. The problem, however, is that
disconnected diagrams are quite difficult to estimate,  so
that additional parameters have to be introduced.

In what follows, we take an alternative viewpoint which
includes a glueball
component in the $\eta'$ wave function, partly to
avoid the parameters of the disconnected diagrams,
partly because there are theoretical reasons why
the $\eta'$ should contain an appreciable admixture
of glue [3], and partly because of arguments [7]
that just singlet-octet mixing is insufficient to explain
the experimental data.

We shall, however, take advantage of the already mentioned
indication [26, 27] that no gluonic component seems to
be present in the $\eta$ to suggest a simpler
mixing scheme than usual, i.e., one in which (i)
only the $\eta$, $\eta'$, and $\eta(1440)$ are mixed, and
(ii) the $\eta$ is a pure two-quark state.
Hereafter, we call the $\eta(1440)$
by its former name $\iota$ for short.

In the past, many different models of mixing between
the $\eta$, the $\eta'$, the $\iota$, and sometimes the
$\eta_{c}$ and the $\eta(1295)$ have been proposed
[7, 30--36].
Although our
model is {\it less}, rather than {\it more}
general than some previous schemes, it has the advantage
that it contains fewer free parameters than other
models, and so has more predictive power.
The simplifications of our
model allow us to fix the mixing parameters using only the
two-photon decay widths of the $\eta$ and $\eta'$
and to make many testable predictions.
It will be for experiment to decide whether our
model is adequate.

There is probably more than one flavorless
pseudoscalar state with mass
in the region 1400 to 1500 MeV. Although
the Particle Data Group [2] lists in its summary table
only the $\eta(1440)$ in this region, the
meson full listing contains a discussion which
points out that the $\eta(1440)$ probably consists of
two states with quantum numbers $I^G=0^+$, $J^{PC}=
0^{-+}$. One of these states may be
at about 1410 MeV, and the other may be at
about 1490 MeV, as can be seen from the full listings
of the Particle Data Group [2];
see also M. G. Rath {\it et al.}\ [37].
(We note, however, that one experiment
[38] finds some
evidence for two pseudoscalar states around 1410 MeV.)
Of the states around 1410 and 1490, one is probably
a quark-antiquark excitation of the $\eta$ and the other is
a good candidate to have a large admixture of glue.
The predominantly glueball state
whichever it is, is the one we call $\iota$.
We keep the discussion rather general with respect
to the iota mass,
presenting results for the different mass values 1410,
1440, and 1480 MeV.

We show below that our
mixing scheme, with no glue in the $\eta$ but with
a gluonic component in the $\eta'$ and $\iota$,
is able to account for many experimental observations.
The success of our scheme does not
mean that the contributions of disconnected diagrams in
$J/\psi$ decays are negligible, but
only that they do not have the large effect
required by Refs. [26,29].
A possible, quite recent, further indication of
unusual behavior of the $\eta'$ comes from a
comparison of the $\eta'$ production rate
with the predictions of the HERWIG [39]
and JETSET [40] Monte Carlo programs, as carried out by
ALEPH [41].
Numerical simulations
with purely quarkonic components for both the $\eta$
and the $\eta'$, overestimate this rate.
A similar situation
occurs in the analyses of
the ARGUS data [42] on $\eta'$ production in
the nonresonant continuum of $e^+ e^-$
around $\sqrt{s} \approx 10 GeV$ and
of the decays of the
$\Upsilon$ resonances. Here again the LUND [43] and the
UCLA [44] models considerably overestimate
the experimental rates.

We introduce the notation
$$| \eta_8 \rangle  ={ 1 \over \sqrt{6}}
|u \bar{u}+ d \bar{d} - 2 s \bar{s}\rangle, \quad
|\eta_1 \rangle= {1
\over \sqrt{3}} | u \bar{u}+d \bar{d}+ s \bar{s}\rangle,
\eqno(1) $$
$$|q\bar q \rangle=
{1\over \sqrt 2}|u\bar u + d\bar d\rangle, \eqno(2)$$
and a pure glueball state $|G\rangle$.
We then consider the following admixtures:
$$ \eqalignno{ | \eta \rangle=  &
cos \alpha |\eta_{8}\rangle +
 \sin \alpha | \eta_1 \rangle \cr
 =& a_{11} | \eta_8 \rangle + a_{12} | \eta_1 \rangle \cr
 =& X_{\eta} | q \bar q \rangle +
Y_{\eta} | s \bar s \rangle \cr
| \eta' \rangle =&-\cos \beta \sin \alpha |
\eta_{8}\rangle +\cos \beta
 \cos \alpha | \eta_1 \rangle + \sin \beta | G \rangle \cr
=& a_{21} | \eta_8 \rangle + a_{22} | \eta_1 \rangle +
a_{23} | G \rangle & (3) \cr
 =& X_{\eta'} | q \bar q \rangle + Y_{\eta'}
| s \bar s \rangle + Z_{\eta'} | G \rangle \cr
| \iota \rangle = & \sin \beta \sin \alpha
| \eta_{8}\rangle -\sin \beta
 \cos \alpha | \eta_1 \rangle + \cos \beta | G\rangle  \cr
 =& a_{31} | \eta_8 \rangle + a_{32} | \eta_1 \rangle +
a_{33} | G \rangle \cr
 =& X_{\iota} | q \bar q \rangle +
Y_{\iota} | s \bar s \rangle + Z_{\iota} |G \rangle,\cr}$$
where the coefficients of the various
states are constant parameters to be determined.
Hereafter, we use the symbols $X_P$, $Y_P$, and $Z_P$
to refer to the coefficients of $|q\bar q\rangle$,
$|s\bar s\rangle$, and $|G\rangle$ respectively, when
referring to any pseudoscalar meson $P$.
Our notation follows that of Caruso {\it et al.}\ [7].
In the case in which the glueball components of $\eta$
and $\eta'$ are both absent, the angle $\alpha$ above
is related to the angle $\theta_P$ of the Particle
Data Group [2] by $\alpha=-\theta_P$.

 In order to fix the two mixing angles $\alpha$ and $\beta$
(the third mixing angle
that one normally has in  three-particle mixing is absent
because of our choice of a
mixing scheme), we use the experimental data:
$$\eqalign{ &R_{\eta} \equiv
 { \Gamma (\eta \rightarrow \gamma \gamma) \over
\Gamma (\pi^0 \rightarrow \gamma \gamma)} =
{1 \over 3 } ( {m_{\eta} \over m_{\pi^0}
} )^3 ({f_{\pi} \over f_{\eta}})^2 (a_{11}+2
\sqrt{2} a_{12})^2 = 59.8, \cr
&R_{\eta'} \equiv { \Gamma (\eta' \rightarrow \gamma
\gamma) \over \Gamma (\pi^0 \rightarrow \gamma \gamma) } =
{1 \over 3 } ( {m_{\eta'} \over m_{\pi^0}
} )^3 ({f_{\pi} \over f_{\eta'}})^2 (a_{21}+
2 \sqrt{2} a_{22})^2 = 555,} \eqno(4)$$
where $(  m_i / m_{\pi^0} )^3$ is a kinematical
(phase space) factor and the $f_i$ are decay constants.

In Eqs.\ (4), $R_{\eta}$ and $R_{\eta'}$ depend on
$\alpha$ and $\beta$ through the quantities $a_{ij}$.
Following Caruso {\it et al.}\ [7], we introduce the quantity
$\tilde R_{i}$:
$$\tilde R_{i} \equiv
3 R_i ( m_{\pi^0} / m_i )^3 (f_{i} / f_{\pi})^2,
\quad i = \eta,\ \eta',\ \iota.  \eqno(5)$$
Then we can get $\alpha$ by eliminating
the dependence on $\beta$ from Eqs. (4), obtaining
$$\sin ( \alpha+ \arcsin { 1 \over 3} ) = \pm
{1 \over 3 }  \tilde R_{\eta}^{1 / 2}  \eqno(6)$$
The quantity $\tilde R_{\eta}$ is known, and we get
$\alpha =  13.7^\circ $.
Analogously, we can eliminate
the $\alpha$ dependence from Eqs.\ (4), getting
$$\sin^2 \beta= { 9- \tilde R_{\eta}-\tilde R_{\eta'} \over
9- \tilde R_{\eta}} \eqno(7)$$
Because the left side of Eq.\ (6) is a trigonometric
function and the left side of Eq.\ (7) is a square of
a trigonometric function,  the first must be between
$-1$ and 1 and the second between 0 and 1.
We can use these limits to obtain  a bound on
the ratios $ f_{\eta}/ f_{\pi}$ and
$ f_{\eta'} / f_{\pi}$, given by the inequality [7]:
$${1 \over 3} R_{\eta'} ({m_{\pi_0} \over m_{\eta'}})^3
({ f_{\eta'} \over f_{\pi}})^2 + {1 \over 3}
R_{\eta} ({m_{\pi_0} \over m_{\eta}})^3 ({ f_{\eta}
\over f_{\pi}})^2 \le 1. \eqno(8)$$
Then, using the definition of $R_i$, and
setting either $R_\eta$ or $R_{\eta'}$ equal to zero,
we obtain the following upper bounds
$$ f_{\eta} \le 1.83 f_{\pi}, \quad
 f_{\eta'} \le 1.39 f_{\pi}, \eqno(9)$$
where the the equality sign corresponds to $\beta=0$, namely
to an unmixed glueball.
We get  from Eq.\ (7) $\beta= 30.8^{\circ}$ when we
set $f_{\pi}=f_{\eta}=f_{\eta'}$. This assumption is
justified, both from
their approximate experimental equality [2] and
from theoretical considerations. Theoretically,
the approximate equality
holds either in the Nambu--Jona-Lasinio model [45]
or with a Wess-Zumino
Lagrangian [46] as well as in lattice calculations [47].

We obtain from the previous mixing angles the results:
$$ \eqalign {&X_{\eta}=0.75, \,\, Y_{\eta}=-0.66, \cr
& X_{\eta'}=0.56, \,\, Y_{\eta'}=0.65, \,\,
Z_{\eta'}=0.51, \cr
& X_{\iota}=-0.34, \,\, Y_{\iota}=-0.39, \,\,
Z_{\iota}=0.86.}  \eqno(10)$$
We note that the sign of $Y_P$ depends on
who does the analysis, but the sign is usually irrelevant
because  most experiments determine only the square.
Both signs are found
in the literature, and often only the modulus is given.
The predictions given in Eq.\ (10)
can be compared with experimental data.
The results of
the DM2 collaboration
[26] are (we use
their fit without disconnected diagrams):
$$ X_{\eta}=0.732 \pm 0.039,\,\,
|Y_{\eta}| =0.667 \pm 0.065, $$
$$ X_{\eta'}=0.335 \pm 0.063, \,\,
Y_{\eta'}=0.623 \pm 0.061. \eqno(11)$$
The Mark III  results (Perrier [27]) are:
$$X_{\eta}^2+Y_{\eta}^2=1.1 \pm 0.1, \quad
X_{\eta'}^2+Y_{\eta'}^2=0.65 \pm 0.1. \eqno(12)$$
Haber and Perrier [48] have reanalyzed Mark III data,
obtaining:
$$ X_{\eta}=0.63 \pm 0.05, \,\, Y_{\eta}  =0.80 \pm 0.12 $$
$$ X_{\eta'}=0.36 \pm 0.05, \,\,
Y_{\eta'}=0.69 \pm 0.11. \eqno(13) $$
The Crystal Barrel [49] result is
$$\biggl ({X_\eta ' \over X_\eta} \biggr )^2 =
0.585 \pm 0.008.\eqno(14)$$
Our predictions,
given in Eqs. (10), are in fair agreement with
the experimental results, given in Eqs. (11)--(14).
Incidentally, the Crystal Barrel group, following
DM2 [26],
did their analysis assuming that neither the $\eta$ nor
$\eta'$ contains any glueball content. They obtained
a mixing angle $\alpha=(17.3\pm 1.8)^\circ$, which should
be compared to our result $\alpha=13.7^\circ$.

In our mixing scheme we can also estimate
the so-called box-anomaly contribution to
$\eta ~[\eta'] \rightarrow \gamma \pi^+ \pi^-$.
Following the model of
Refs.\ [50,51], we estimate the
contribution of the box anomaly by considering the
effective amplitude $M_P$, given by
$$M_P = E_P(p_{\pi^+}  k_\gamma, p_{\pi^-}
 k_\gamma) \epsilon_{\alpha \beta \mu \nu}
\epsilon^\alpha_\gamma
k^\beta_\gamma p^\mu_{\pi^+} p^\nu_{\pi^-}, \eqno(15)$$
where $(P=\pi^0,~\eta,~\eta').$
In (15), $k_\gamma$ and $p_{\pi}$ are, respectively,
the four-momenta of the photon and of
the two pions  and $\epsilon$ is the polarization
of the photon.
At low energies and in the approximation in which the decay
constants for the octect and siglet pseudoscalar
states are equal to $f_{\pi}$, the functions
$E_P(p_{\pi^+}  k_\gamma, p_{\pi^-}  k_\gamma)$
reduce to the constants [50]
$$ E_{\eta} = {- e \over 4 \pi^2 \sqrt{3} f_{\pi}^3}
( \cos \alpha + \sqrt{2} \sin \alpha), \eqno(16)$$
$$ E_{\eta'} = {- e \over 4 \pi^2 \sqrt{3} f_{\pi}^3}
( - \sin \alpha + \sqrt{2} \cos \alpha)  \cos \beta.
\eqno(17)$$
With our mixing parameters, we find
$$E_{\eta}=-7 ~({\rm GeV})^{-3},
\quad  E_{\eta'}=-5.3 ~({\rm GeV})^{-3}. \eqno(18)$$
These values are compatible with the results [51]
$$E_{\eta}=-5 \pm 1.5 ~({\rm GeV})^{-3},
\quad E_{\eta'}=-5.1 \pm 0.7
{}~({\rm GeV})^{-3}, \eqno(19)$$
where both the box anomaly term
(15) and the effect of the $\rho$ resonance have been
included as independent contributions to fit the data.
The $\rho$ contribution in Ref. [51]
has been taken into account
using a relativistic Breit-Wigner amplitude where the
width has been parameterized by
$$\Gamma_{\rho}(m)= \Gamma_{\rho}( m_{\rho})  \biggl
[ {q_{\pi}(m) \over
q_{\pi}(m_{\rho})} \biggr ]^3 \biggl
( {m_{\rho} \over m} \biggr )^{\lambda}, \eqno(20)$$
where $q_{\pi}$ is the pion center-of-mass momentum.
It is evaluated with the assumption
that either a physical $\rho$ ($q_{\pi}(m_\rho)$)
or virtual $\rho$ ($q_{\pi}(m)$)  decays.
The parameters $m_{\rho}$, $\Gamma_{\rho}(m_{\rho})$ and
$\lambda$ have been determined from the process
$e^+ e^- \rightarrow \pi^+ \pi^-$ in the $\rho$ region.

We now turn to the the mass of the glueball. The masses of
the pure states in terms of the physical-state masses are:
$$\eqalign{ &m_{\eta_8}=\langle \eta_8 | H |\eta_8 \rangle=
 a_{11}^2 m_\eta+a_{21}^2 m_{\eta'}+ a_{31}^2 m_\iota, \cr
 &m_{\eta_1}=\langle \eta_1 | H | \eta_1 \rangle =
 a_{12}^2 m_\eta+a_{22}^2 m_{\eta'}+ a_{32}^2 m_\iota, \cr
 &m_{G}=\langle G | H | G \rangle =
 a_{23}^2 m_{\eta'}+ a_{33}^2 m_\iota, }\eqno(21) $$
where $H$ is the Hamiltonian.
Using the last of Eqs.~(21) and the
mixing parameters (10) we get
for $m_\iota=1410$, 1440, 1480 MeV,
$$m_G=1302,\quad 1324,\quad 1354~ {\rm MeV} \eqno(22)$$
 respectively.
These results are compatible with some other
theoretical predictions (see Table I).

Our analysis has allowed us to
derive theoretical expectations for
the relative admixture of $ u \bar u + d \bar d$,
$ s \bar s$ and $G$ for $\eta$, $\eta'$ and $\iota$.
Therefore, an accurate measurement of the mass of the
$\iota$ gives us the
mass of the glueball state $|G\rangle$, which can be
compared to calculations with lattice gauge theory or
phenomenological models.

\bigskip

\centerline{\bf C. Comparison of the quark composition
of the $\eta$ and $\eta'$ with data}

\bigskip

In order to check our mixing scheme we compare the ratios
between some decay widths of the $\eta$ and of the
$\eta'$ evaluated in our model with the experimental data.
This comparison also enables us to tell which
future measurements will be useful.

In the following, we use known decay widths to evaluate
unknown widths,
taking into account the appropriate kinematic factors
and using SU(3) calculations of the amplitudes [52,53].
We do this by selecting among possible decay modes
in our mixing scheme those decays with only one
channel allowed by
the OZI (Okubo--Zweig--Iizuka) rule,
which requires that hadronic reactions are suppressed
when the corresponding quark diagrams have disconnected
quark lines from the initial hadrons to
the final ones [54].
To this end, we choose reactions
involving at least one particle which is purely $q \bar q$
or $s \bar s$. In this case there will appear factors
proportional to
certain parameters of the mixing scheme for which
we can use our previous results (10).

Using the known branching fraction
$B(\phi \rightarrow \eta \gamma )= 0.0128 \pm 0.0006 $, we
obtain:
$$ B(\phi \rightarrow \eta' \gamma ) = \Biggl [{
 m_{\phi}^2-m_{\eta'}^2
\over  m_{\phi}^2-m_{\eta}^2 }\Biggr ]^3
\biggl ({ Y_{\eta'} \over Y_{\eta}} \biggr )^2
B(\phi \rightarrow \eta \gamma ) =
( 5.6 \pm 0.3 ) \times 10^{-5}, \eqno(23)$$
a result which is compatible with the experimental bound
$B(\phi \rightarrow \eta' \gamma ) < 4.1 \times 10^{-4}$.

Similarly, from the observed $\omega$ decay branching
fraction
$B (\omega \rightarrow \pi^{0} \gamma )=
(0.085 \pm 0.005)  $
we obtain the result
$$ \Gamma (\rho \rightarrow \eta \gamma ) =
\Biggl [{ ( m_{\rho}^2-m_{\eta}^2)
\over ( m_{\omega}^2-m_{\pi}^2)} {m_\omega \over m_\rho}
\Biggr ]^3 X_{\eta}^2
\Gamma (\omega \rightarrow \pi^0 \gamma ) =
( 5.0 \pm 0.3 ) \times 10^{-2} {\rm MeV}, \eqno(24)$$
in good agreement with the experimental result
$\Gamma (\rho \rightarrow \eta \gamma ) =
 ( 5.7 \pm 1.4 ) \times 10^{-2}$ MeV.
Furthermore, we have
$$ \Gamma (\phi \rightarrow \eta \gamma ) =
 \Biggl[ {( m_{\phi}^2-m_{\eta}^2)
\over ( m_{\omega}^2-m_{\pi}^2 )}
{m_\omega \over m_\phi}\Biggr ]^3
{4 \over 9} Y_{\eta}^2 ({m_{u} \over m_s})^2
\Gamma (\omega \rightarrow \pi^0 \gamma ), \eqno(25)$$
where we have taken into account $SU_F(3)$ ($F$ for flavor)
symmetry breaking [53],
by assuming the strange-quark magnetic moment
to be smaller than the $u$-quark
moment by a factor $m_u/m_s$.
If we assume $m_u / m_s = 3 /5$, then
$ \Gamma (\phi \rightarrow \eta \gamma ) = 0.044$
MeV. This assumption gives sort of an upper bound on
$ \Gamma (\phi \rightarrow \eta \gamma ) $
because at the $\phi$ energy the quark masses should
be between the constituent and the current masses [55].
The experimental result is
$\Gamma (\phi \rightarrow \eta \gamma ) = 0.057 \pm 0.027$
MeV, compatible with our bound.

Next, we have
$$ \Gamma (\eta' \rightarrow \rho \gamma ) = 3  \Biggl
[ { (m_{\eta'}^2-m_{\rho}^2)
\over ( m_{\omega}^2-m_{\pi}^2 )} \
{m_\omega \over m_{\eta'}} \Biggr ]^3 X_{\eta'}^2
\Gamma (\omega \rightarrow \pi^0 \gamma ) = 0.062
\pm 0.004  \ {\rm MeV}, \eqno(26)$$
again, compatible with the experimental finding
$\Gamma (\eta' \rightarrow \rho \gamma )
 = 0.059 \pm 0.003$  MeV.

We can calculate another set of branching ratios in the
charmed sector, assuming the
dominance of spectator quark diagrams [53].
Because of this assumption, the following
results can at the best be approximations to the
real situation. However, we should at least
get the right order of magnitude.

We parameterize the partial width for a decay into final
states with orbital angular momentum $l$ as [53]
$$ \Gamma \approx \tilde \Gamma (k/2M)^{2l+1},  \eqno(27)$$
where $\tilde \Gamma$ is the partial width with
kinematic factors taken out, k denotes the
center-of-mass three-momentum, and M is the
mass of the decaying particle.
For an S wave, using the experimental value
 $B(D_s^+ \rightarrow \eta \pi^+ )= 0.015 \pm 0.004 $,
we get from Eq.\ (27):
$$ \eqalign{ &B(D_s^+ \rightarrow \eta' \pi^+ )\cr & =
\Biggl[{ ( m_{D_s^+}^2-
(m_{\eta'}+m_\pi)^2)  ( m_{D_s^+}^2-
(m_{\eta'}-m_\pi)^2)
\over ( m_{D_s^+}^2-
(m_{\eta}+m_\pi)^2)  ( m_{D_s^+}^2-
(m_{\eta}-m_\pi)^2)} \Biggr ]
^{1 \over 2 } ({ Y_{\eta'} \over Y_{\eta}})^2
B(D_s^+ \rightarrow \eta \pi^+ ) \cr &= ( 1.2 \pm 0.3 ) \%,}
\eqno(28)$$
to be compared with the experimental result
$(3.7 \pm 1.2 ) \%$.
In the same hypothesis we get the predictions:
$$B(D_s^+ \rightarrow \iota \pi^+ ) =
( 2.4 \pm 0.6 ) \times 10^{- 3}\quad {\rm or}\
B(D_s^+ \rightarrow \iota \pi^+ ) =
( 2.6 \pm 0.7 ) \times 10^{- 3} \eqno(29)$$
respectively for $m_\iota = 1480$ MeV and
$m_\iota = 1410$ MeV.

Another interesting set of comparisons with experimental
data can be obtained from the decays of the $J/\psi$
into a vector and a
pseudoscalar meson. The phase space factor is given by
the modulus of the ratio of the center-of-mass
three-momenta of the final particles raised to third power.
We compare processes which are expressed in
terms of the same amplitudes [26, 48, 56]
and we work in the approximation
in which the contributions of doubly disconnected
(or doubly OZI suppressed) diagrams are neglected. We get
$$ B(J/\psi \rightarrow \omega \eta)  =
X_{\eta}^2 \biggl \vert
{k_{\eta \omega} \over k_{\rho \pi}} \biggr \vert^3
B(J/\psi \rightarrow \rho^0 \pi^0) =
( 2.09 \pm 0.25 ) \times 10^{-3}, \eqno(30)$$
in fair agreement with the
observed value $( 1.58 \pm 0.16 ) \times 10^{-3}$.
We also obtain
$$ B(J/\psi \rightarrow \omega \eta')  =
X_{\eta'}^2 \biggl \vert
{k_{\eta' \omega} \over k_{\rho \pi}} \biggr \vert^3
B(J/\psi \rightarrow \rho^0 \pi^0) =
( 0.89 \pm 0.11 ) \times 10^{-3}, \eqno(31)$$
which grossly overestimates the observed value
$( 1.67 \pm 0.25 ) \times 10^{-4}.$
It should, however, be pointed out that
in the case of $J/\psi \rightarrow \omega \eta'$, the
procedure used (i.e. the neglect of the doubly OZI
suppressed diagrams) leads us to overestimate
the branching ratio (31).
As it turns out, in the presence of a gluonic component
of the $\eta'$, one finds that the
neglected diagrams would in fact lower our result;
see Jousset {\it et al.}\ [26].

For the ratio of $J/\psi$ decays into $\phi\eta$
and $\phi\eta'$ we find
$$ {B(J/\psi \rightarrow \phi \eta) \over
B(J/\psi \rightarrow \phi \eta')} =
\biggl ({ Y_{\eta} \over Y_{\eta'}} \biggr )^2 \biggl \vert
{k_{\eta \phi} \over k_{\eta' \phi}} \biggr \vert^3 = 1.42,
\eqno(32) $$
which approximately
agrees with the observed value $1.97 \pm 0.45$.
The already mentioned neglect of doubly OZI suppressed
diagrams appears
not to be as important as in the previous case.
Nevertheless, including such  diagrams somewhat
improves the agreement with the data as this
leads to an increase of the ratio (32).  We also get
$$ B(J/\psi \rightarrow \rho^0 \eta)  =
X_{\eta}^2 \biggl \vert
{k_{\eta \rho} \over k_{\omega \pi}} \biggr \vert^3
B(J/\psi \rightarrow \omega \pi^0) =
( 2.12 \pm 0.30 ) \times 10^{-4}, \eqno(33)$$
in good agreement with the observed value
$( 1.93 \pm 0.23 ) \times 10^{-4}$.
Furthermore, we get
$$ B(J/\psi \rightarrow \rho^0 \eta')  =
X_{\eta'}^2 \biggl \vert
{k_{\eta' \rho} \over k_{\omega \pi}} \biggr \vert^3
B(J/\psi \rightarrow \omega \pi^0) =
( 0.91 \pm 0.14 ) \times 10^{-4}, \eqno(34)$$
in good agreement with the observed value
$( 1.05 \pm 0.18 ) \times 10^{-4}$.

Using analogs
of Eqs.\ (30), (32), and (33), we can now make predictions
for the branching ratios for
$J/\psi \rightarrow \iota + V$ where $V$ is a vector meson.
Using $ m_{\iota}=1410$ MeV, we get
$$ B(J/\psi \rightarrow \omega \iota)  = \biggl (
{X_{\iota} \over X_{\eta}}\biggr ) ^2 \biggl \vert
{k_{\iota \omega} \over k_{\eta \omega}} \biggr \vert^3
B(J/\psi \rightarrow \eta \omega)  =
( 1.5 \pm 0.2 ) \times 10^{-4} \eqno(35)$$
and
$$ B(J/\psi \rightarrow \rho^0 \iota)  = \biggl (
{X_{\iota} \over X_{\eta}}\biggr ) ^2 \biggl \vert
{k_{\iota \rho} \over k_{\eta \rho}} \biggr \vert^3
B(J/\psi \rightarrow \eta \rho^0) =
( 1.9 \pm 0.2 ) \times 10^{-5}. \eqno(36) $$
We have two expressions for
$ B(J/\psi \rightarrow \phi \iota)$, depending on whether
we use the decay width into $\phi\eta$ or $\phi\eta'$
as input:
$$ B(J/\psi \rightarrow \phi \iota)
 = \biggl ({ Y_{\iota} \over Y_{\eta}}
\biggr )^2 \biggl \vert
{k_{\iota \phi} \over k_{\eta \phi}} \biggr \vert^3
B(J/\psi \rightarrow \phi \eta)=
( 0.86 \pm 0.09 ) \times 10^{-4} \eqno(37)$$
and
$$ B(J/\psi \rightarrow \phi \iota)
 = \biggl ({ Y_{\iota} \over Y_{\eta'}}
\biggr )^2 \biggl \vert
{k_{\iota \phi} \over k_{\eta' \phi}} \biggr \vert^3
B(J/\psi \rightarrow \phi \eta')=
( 6.0 \pm 0.7 ) \times 10^{-5}. \eqno(38)$$
The above results are only marginally dependent on
which value one uses for the $\iota$ mass. In fact,
using $ m_{\iota}=1480$ MeV we obtain
$$ B(J/\psi \rightarrow \omega \iota) =
( 1.4 \pm 0.1 ) \times 10^{-4}, \eqno(39) $$
and
$$ B(J/\psi \rightarrow \rho^0 \iota) =
( 1.7 \pm 0.2 ) \times 10^{-5}. \eqno(40) $$
while the results for$ B(J/\psi \rightarrow \phi \iota)$
become:
$$ B(J/\psi \rightarrow \phi \iota) =
( 0.71 \pm 0.08 ) \times 10^{-4} \eqno(41) $$
and
$$ B(J/\psi \rightarrow \phi \iota) =
( 5.2 \pm 0.6 ) \times 10^{-5}. \eqno(42) $$
The only present experimental information, the
upper limit [2]
$ B(J/\psi \rightarrow \phi \iota) < 2.5 \times 10^{-4}$,
is consistent with our values given in Eqs.\
(37, 38, 41, 42).

Next, we give predictions for various $\iota$ decays.
Using the fact that the amplitude  for a pseudoscalar meson
$P$ decaying into
$ \rho^0 \gamma$ is
proportional to $X_P$
and that the kinematical factor is proportional to
$ [ m_P-(m_\rho^2 / m_P )]^3$
we obtain:
 $$\eqalign { \Gamma( \iota \rightarrow \rho^0 \gamma) = &
\Biggl [ { m_\iota-m_{\rho}^2/m_\iota
\over m_{\eta'}-m_{\rho}^2/
m_{\eta'}} \Biggr ]^3 \tan ^2 \beta
\Gamma( \eta' \rightarrow \rho^0 \gamma) \cr
= &  0.51 \pm 0.02,
\, 0.58 \pm 0.03, \, 0.67 \pm 0.03\ {\rm MeV}} \eqno(43)$$
according to whether $m_\iota = 1410, \, 1440,\ or\ 1480$ MeV
respectively.  Similarly, from
$$ \Gamma( \iota \rightarrow \phi \gamma) =
\Biggl [ { m_\iota-m_{\phi}^2/m_\iota
\over m_{\phi}-m_{\eta'}^2/
m_{\phi}} \Biggr ]^3 \biggl ( { Y_\iota \over Y_{\eta'} }
\biggr)^2  \Gamma( \phi \rightarrow \eta' \gamma),
\eqno(44) $$
we get the upper bound
$$ \Gamma( \iota \rightarrow \phi \gamma) < 0.12,\
0.14,\ 0.18\ {\rm MeV}$$
for $m_\iota = 1410, \, 1440,\ and \ 1480$ MeV respectively.
Using the analog of Eqs.\ (4) for the $\iota$, we get
$$\eqalign{ \Gamma( \iota \rightarrow \gamma \gamma) =  & 3
( { m_\iota \over m_{\pi^0}} )^3
\sin ^2 \beta \cos^2 (\alpha+
\arcsin (1/3)) ( {f_\pi \over f_\iota} )^2 \
\Gamma( \pi^0 \rightarrow \gamma \gamma)\cr
= & 4.86  ( {f_\pi \over f_\iota} )^2,
\, 5.17  ( {f_\pi \over f_\iota} )^2,
\, 5.62  ( {f_\pi \over f_\iota} )^2\ {\rm keV}} \eqno(45)$$
respectively for $m_\iota = 1410, \, 1440,\, 1480$ MeV.

The next thing we can do is to derive the
branching ratio for
$J/\psi \rightarrow\iota \gamma$,
assuming that the decay goes  principally through the
gluonic component. The branching ratio for the
radiative decay of the $J /\psi$ into the $\eta$
[$B(J/\psi \rightarrow
\eta \gamma) \approx 8.6 \times 10^{-4}$]
gives an estimate of the
decay into a pure quarkonic state and
consequently  gives us an idea of the error made
by treating the $\iota $ as a pure gluonic state.
With this assumption we find (for the three
values of $m_\iota $ specified above):
$$ \eqalign{ &B(J/\psi \rightarrow
\iota \gamma)= \Biggl [ { m_{J/\psi}^2-m_{\iota}^2 \over
m_{J/\psi}^2-m_{\eta'}^2} \Biggr ]^{3} \cot^2 \beta \
 B(J/\psi \rightarrow\eta' \gamma)  \cr
= &( 8.3 \pm 0.6 )\times 10^{-3}, \,
 ( 8.0 \pm 0.6 )\times 10^{-3}, \,
( 7.6 \pm 0.6 )\times 10^{-3}. }\eqno(46)$$
The experimental branching ratio for the $J/\psi$
to decay radiatively into the $\eta(1440)$ is [2]
$( 2.4 \pm 0.4 ) \times 10^{-3}$, a factor 3 smaller
than our estimate.
The above result, if taken at face value,
would suggest some serious discrepancies with the
assumption leading to Eq.\ (46),
i.e. the assumption that the decays
$J/\psi \rightarrow \iota
\gamma$ and  $J/\psi \rightarrow \eta' \gamma$ go
mainly through the gluonic component.  On the other hand,
as we have already mentioned, there are probably
several resonances coexisting in the 1400--1500 MeV
region. The above discrepancy may thus
be rather an indication of serious
problems of experimental resolution.

The assumption
that the $J/\psi$ radiative decays go mainly through
the gluonic component could also be tested in
$\psi (2S)$ and $\Upsilon$
decays. From the analog of Eq.\ (46) we get
$$ \eqalign{ B(\psi (2S) \rightarrow \iota \gamma)=
2.12  B(\psi (2S) \rightarrow \eta' \gamma)  }
\eqno(47)$$
and
$$ \eqalign{ B(\Upsilon \rightarrow
\iota \gamma)=
2.71  B(\Upsilon \rightarrow \eta' \gamma)  }.
\eqno(48)$$
Only the following experimental
upper bounds are known at present:
$$B(\psi (2S) \rightarrow \eta' \gamma)
< 1.1 \times 10^{-3}, \eqno(49)$$
$$B(\psi (2S) \rightarrow \iota \gamma) \
B( \iota \rightarrow \pi K \bar K \gamma )
< 1.2 \times 10^{-4},\eqno(50)$$
$$B(\Upsilon \rightarrow \eta' \gamma) < 1.3 \times 10^{-3},
\eqno(51)$$
$$B(\Upsilon \rightarrow \iota \gamma) < 8.2 \times 10^{-5},
\eqno(52)$$
which are inadequate to test the
validity of the assumption leading to Eqs.\ (47) and (48).
New and better data are needed to settle the issue.

Next we give an estimate of the cross section
$ \sigma(e^+ e^- \rightarrow e^+ e^- \iota)$ in the
equivalent photon approximation, namely the process in which
the pseudoscalar particle (the $\iota$ in this case),
is produced by the interaction of the
two virtual photons emitted by electron and
positron (see Fig. 1). The idea is pretty old [57],
and the resulting formula for $\sigma$ is [7, 58]
$$ \sigma(e^+ e^- \rightarrow e^+ e^- \iota)=
16 \alpha^2 \biggl ({ 1 \over m_{\iota}^3} \biggr)
\biggl [ \ln( { E \over m_e})-{1 \over 2} \biggr ]^2
f({m_\iota \over 2E}) \
\Gamma(\iota \rightarrow \gamma \gamma), \eqno(53)$$
where $E$ is the beam energy and $f(x)$ is given by
$$f(x)=(2+x^2)^2  \ln(1/x)-(1-x^2)(3+x^2). \eqno(54)$$
Using our previous estimates of
$\Gamma(\iota \rightarrow \gamma \gamma)$, we obtain
$\sigma$
as a function of $f_\pi / f_\iota$ and for various energies:
$$\  E= 2 \ {\rm GeV}, ~~ \sigma= (68,\ 66,\ 64)
 \ (f_\pi / f_\iota )^2\ {\rm pb},$$
$$E= 3 \ {\rm GeV}, ~~  \sigma= (124,\ 121,\ 118)
 \ (f_\pi / f_\iota )^2 \ {\rm pb},$$
$$E= 10 \ {\rm GeV},  ~~ \sigma= (388,\ 383,\ 378)
 \ (f_\pi / f_\iota )^2 \ {\rm pb}, \eqno(55)$$
$$E= 50 \ {\rm GeV}, ~~ \sigma= (976,\ 969,\ 963)
 \ (f_\pi / f_\iota )^2 \ {\rm pb}  $$
for $\iota$ masses 1410, 1440, and 1480 MeV.
The ratios are independent of the uncertainties related to
$ f_\pi / f_\iota $. We get
$$\eqalign{ &{\sigma(E=2\ {\rm GeV}) \over
\sigma(E=3\ {\rm GeV})}= 0.548,\ 0.546, \ 0.542 \cr
&{\sigma(E=2\ {\rm GeV}) \over \sigma(E=10\ {\rm GeV})}=
0.175, \ 0.172,\ 0.169 \cr
&{\sigma(E=2\ {\rm GeV}) \over \sigma(E=50 \ {\rm GeV})}=
0.069 \ 0.068,\ 0.066\cr
&{\sigma(E=3\ {\rm GeV}) \over \sigma(E=10\ {\rm GeV})}=
0.320 \ 0.316,\ 0.312\cr
&{\sigma(E=3\ {\rm GeV}) \over \sigma(E=50\ {\rm GeV})}=
0.127 \ 0.125,\ 0.123\cr
&{\sigma(E=10\ {\rm GeV}) \over \sigma(E=50\ {\rm GeV})}=
0.400 \ 0.396,\ 0.393.
\cr} \eqno(56)$$
Good data on  $\gamma \gamma$ decay
and on the cross section (53)
would help us obtain a good value of $ f_\pi / f_\iota $.
\bigskip

In summary, we find that
our mixing scheme, with a glueball
component in the $\eta'$ but not in the
$\eta$, gives reasonable agreement
with the experiments in
most cases.
The agreement is poorest where the theoretical assumptions
are the most questionable (dominance of spectator diagrams,
gluonic dominance of some decays, etc.).
Further investigations will clarify how well this
simple scheme can explain the pseudoscalar
sector. Before considering more complex
mixing models it makes good sense to test
further
the scheme we have described.
\bigskip

\centerline{III. THE SCALAR SECTOR}
\bigskip
\centerline{\bf A. Problems with the $^3 P_0$ nonet}
\bigskip

Of the $L=1$  meson nonets [mixed octets and singlets
of flavor $SU(3)$],
the scalar, with spin, parity, and charge-conjugation
$J^{PC}=0^{++}$ ($^3 P_0$ states in the quark model)
is the poorest known.
Here, the value of
$C$ refers only to the self-conjugate members.
The $0^{++}$ resonances known at present [2]
are listed in Table II. As we can see from this table,
there are quite a few candidates for members of the
scalar nonet. The difficulty is in the interpretation.

The two states $a_0(980)$ and $f_0(975)$ are especially
difficult to interpret. From here on, we
denote these states in the main text
simply by $a_0$ and $f_0$.
Let us consider some of the difficulties.
Godfrey and Isgur [24] have calculated the masses
of the $P$-wave mesons in their potential model. They
find rather good agreement with the observed masses
of the axial vector and tensor mesons, but not with
the scalars.
These authors suggest that the scalar states composed
of $u$ and $d$ quarks, {\it i.e.}, the $^3 P_0$ states,
have masses around 1090 MeV, over 100 MeV larger than
the masses of the $a_0$ and the $f_0$, but
about 200 MeV smaller than the mass
of the other possible candidate for the role of $0^{++}$
isovector: the $a_0(1320)$.
Thus, the low masses of the $f_0$ and the $a_0$
give us reason to doubt their
being ordinary quarkonium states.
There is an additional problem if the $f_0$
is assumed to be a quarkonium state: the near
degeneracy in mass with the isovector state
$a_0$ seems to require the $f_0$ to be dominantly a
$u \bar u + d \bar d$ meson, while the strong branching
ratio into $ K \bar K$ suggests a large strange component.

Another, admittedly qualitative, argument
which causes us to doubt that the $a_0$
and the $f_0$ are the isovector and isoscalar members
(respectively) of the scalar nonet is the following. One
expects the mass splittings between corresponding members
of the different nonets to be
comparable. Indeed, this is very roughly the case
for the corresponding members of the other L=1 nonets:
$^1 P_0$, $^3 P_1$ and $^3 P_2$.
In contrast, the mass splitting of the $a_0$ and of the
$f_0$ from their corresponding (isovector and
isoscalar) partners in the other nonets is
much larger than the splitting between the $K^*_0(1430)$
(which is likely the strange meson of
the $^3 P_0$ scalar nonet) and the
corresponding $K$ states of the other nonets.


Many different models have been proposed suggesting that
the $a_0$ and the $f_0$ might be:
i) $q \bar q q \bar q$ states [59, 60];
ii) $K \bar K$ molecules [60, 61];
iii) members of a quarkonium nonet mixed with a
$q \bar q q \bar q$ state [62],
or, in the case of the isoscalar state,
with a glueball [60].
The $f_0$ has been interpreted as a dilaton by
Halyu [63].
It seems unlikely that the $a_0$ and the $f_0$
are hybrids, as their masses are much
lower than those  usually assigned to hybrids [13, 18, 22].
On the other hand, the $f_0(1240)$ and the $a_0(1320)$ might
be hybrids [18, 22], as their masses are
considerably higher than those of the $a_0$
and the $f_0$.
Another suggestion connects the
phenomenology of the lowest scalar particles
with the excitation of the QCD vacuum [64]. In this case
a special role is advocated for the light-quark (u,d)
condensate.

Still other $0^{++}$ states do not have a clear
interpretation.  Because of their peculiar decay patterns, a
predominantly gluonic component has been attributed to both
the $f_0(1590)$ [65] and the $f_0(1710)$ [66].
These states might arise from the mixing of a glueball
with an $s \bar s$ state [67].
Also, Alde {\it et al.}\
[68] have suggested that the $f_0(1400)$ may have
a gluonic component.

Predictions of the mass of a scalar glueball
spread over a large interval [9, 11, 13--19, 63, 69--74],
as we show in Table III. Recent lattice [11, 70]
and flux tube  calculations [18]
put the scalar glueball mass around 1500 MeV
(not far from the $f_0(1590)$).
In principle, the calculations using lattice QCD
should be the most reliable. However, the
lattice calculations are made in the valence (or
quenched) approximation, and it is not obvious that
this approximation  is a good one.
Besides that,
calculations in nonrelativistic potential models [15, 21],
bag models [13, 19]
and other lattice simulations [9, 16, 73]
indicate that the scalar glueball mass
is under or around 1000 MeV.  Also, in  a
Bethe-Salpeter equation approach,
Bhatnagar and  Mitra [22] do not find a consistent
solution for the scalar glueball, but find a
tensor glueball mass satisfying
$1200 MeV \lsim m_{2^{++}} \lsim 1600 MeV$.
We have to conclude that, so far, theorists have
not been able to calculate the scalar glueball mass
with any reliability.

A low-mass predominantly glueball state might be
identified with a broad scalar
resonance around 750 MeV. Recent evidence for this
resonance has been given by Svec {\it et al.}\ [75], but the
state should not be regarded as well established.
In this case an excited state should be in
the region of the $f_0(1590)$ and of the $f_0(1710)$.



We mentioned previously that the $a_0$ and $f_0$
may be $K \bar K$ molecules. This
idea has been suggested in order
to explain the approximate mass degeneracy  of the
$a_0$ and $f_0$ and the fact that they are
so near to the $ K \bar K$ threshold.
Also, the assumption that these mesons are quark-antiquark
pairs leads to difficulties.  For example,
Close {\it et al.}\ [76],
interpreting the $f_0$
state as consisting only of $u$ and $d$ quarks, find
$${ \Gamma(f_0 \rightarrow \pi \pi ) \over
 \Gamma(a_0 \rightarrow \pi \eta )} \approx 4, \eqno(57)$$
in contrast with the observed rate 0.6.
 Furthermore, the total widths calculated for
light-quark mesons
in a potential model [24]
are respectively $\approx 1000$
MeV for the $f_0$ and 400 MeV for the $a_0$,
compared to the observed widths of $ 47 \pm 9$ MeV and
$57 \pm 11$ MeV.  In  a flux tube model [77]
the value
$$ \Gamma(f_0 \rightarrow \pi \pi ) \approx 400
\ {\rm MeV} \eqno(58)$$
has been also obtained, compared to the observed
value 37 MeV, and
$$ \Gamma(a_0 \rightarrow \eta \pi ) \approx 225
\ {\rm MeV}, \eqno(59)$$
compared to the observed  value  of $ \approx 57 MeV$.
In other quarkonic models, however, the results
may be considerably different.
All the above observations might be compatible with
both the $a_0$ and the $f_0$ being
molecular states, arising from their diffuse wave functions.

Another indication that the $a_0$ and $f_0$ might be
$K \bar K$ molecules comes
from phase shift analyses in an approach in which
effective meson-meson potentials are used [78--80].
In these papers a comparison is made
between the data and the predictions obtained
with the potentials.  The result is that molecular bound
states in the appropriate mass range are
expected, or, turning things around, that a molecular
model of these particles is admissible.

Arguments against multiquark interpretations have
also be given.  A recent analysis [81]
of the poles of the amplitude suggests
poles on an unphysical Riemann sheet, a result which
favors a resonance over a bound state
hypothesis. Also, the photoproduction
of the $a_0$ is in agreement with calculations in the
$ q \bar q$ scheme, while a prediction for a
multiquark state has not been given.
Furthermore, for $q \bar q q \bar q$ states,
some mass predictions are larger than 1000 MeV [82].

New experimental data on the ratio of the inclusive
reactions $e^+e^- \rightarrow Z_0 \rightarrow f_2(1270) X$
and
$e^+e^- \rightarrow Z_0 \rightarrow f_0 X$,
where $X$ is anything,
have been given by the Delphi collaboration [41,83], which
finds
$$R_i={\sigma(f_2(1270)) \over \sigma(f_0)}
=3^{+7}_{-1}. \eqno(60).$$
Furthermore, the HRS Collaboration [84] finds
from $e^+ e^-$ annihilation at $\sqrt{s}=29 GeV$
that $ R_i=2 \pm 1$,
while the NA27 Collaboration  [85]
finds $ R_i=4.1 \pm 1.5$
from pp-interactions at $\sqrt{s}=27.5 GeV$.
All these data have large errors,
but are not inconsistent with the value $R_i=5$, predicted
from spin-statistics arguments for the ratio of
the tensor to scalar mesons.

Further evidence against a multiquark interpretation of
the $f_0$ comes from analyzing the
ARGUS data [42] on the
partial widths of the decays of $\Upsilon$ states into
$f_0$ mesons plus anything ($X$).
The partial widths into $f_0$ are in fact comparable with
the $\rho_0(770)$ yields once the different spin
structure of the two resonances is taken into account.

It should also be mentioned that the result
$$ R= {\sigma(\pi^{\pm} p \rightarrow f_0 X) \over
\sigma(K^{\pm} p \rightarrow f_0 X)}=1.66 \pm 0.35
\eqno(61)$$
found by the OMEGA Collaboration [86],
using $\pi$ and $K$ beams of 80 and 140 GeV,
is in agreement with the value  2, predicted
by valence quark counting for a light $q \bar q$ state.
(The value is reduced somewhat by
contributions from sea quarks and gluons.)
The ratio R is expected
to fall further  for a state rich in strange quarks
(like a $K \bar K$ molecule). In fact, the process $\pi p
\rightarrow s \bar s X$
is a peripheral one and
involves at least two hard gluons in order
to produce a colorless $s \bar s$ state, while
$K p \rightarrow s \bar s X$ requires only one hard gluon:
in this case, one expects, therefore, in a first
approximation $ R \sim \alpha_s$.

Close {\it et al.}\ [76]
have suggested that a measurement of
the ratio of the radiative $ \phi $ decays in these two
states (see Table 4) might clarify the situation.
However, the $f_0$ may be  far from being
ideally mixed  and might have a big gluonic content,
complicating the situation. In addition, the prediction
given by these authors for the case of a large gluonic
component of the $f_0$ is quite model dependent.

\bigskip

\centerline{\bf B. Phenomenology of the scalar mesons }
\bigskip
In the following, we
suggest some
possible ways to investigate the properties of the
scalar mesons, especially the flavorless ones.
We propose measurements which we believe
can be carried out in the
next experiments (AGS P852, SuperLear, DA$\Phi$NE, etc.)
and can help to discriminate between the
large number of hypotheses about the nature of the
scalars.

Interesting information
about the composition of the low-lying scalar resonances
comes from the ratio [87],
$${ \Gamma(a_0 \rightarrow \eta \pi ) \over
 \Gamma(K_0^*(1430) \rightarrow K \pi )} =
\Biggl [ {m_{a_0}^2-m_\eta^2 \over m_{K^*_0}-m_\pi^2}
\Biggr ]^2 {m_{K^*_0}^2 \over m_{a_0}^2 }
{ f_{K}^2 \over f_\pi^2 } { k_{a_0} \over
k_{K_0^*}}{2 \over 3} \biggl [ \sqrt {2 \over 3} a_{11}
 + {2 \over \sqrt3} a_{12} \biggr ] ^2 \eqno(62)$$
where $a_{11}$ and $a_{12}$ are the mixing parameters
defined previously for the $\eta$ and
$f_K /f_\pi = 1.17$.  The above ratio is
obtained in the PCAC hypothesis for a quarkonic state
[87].
In the scheme proposed here,
we get 0.12 for the r.h.s. of Eq.\ (62).
This should be compared with the value of
0.09 in the model of Caruso {\it et al.}\ [7] and
0.11 in the model of Teshina and Oneda [87].
Furthermore, we find  0.072 in the limiting case
when $\eta=\eta_8$ and  0.144 when $\eta= \eta_1$.

On the other hand, as a first rough approximation,
we can assume
$$\Gamma(a_0 \rightarrow \eta \pi )
\approx \Gamma( a_0 ) = 57 \pm 11\ {\rm MeV},
\eqno(63)$$
where $\Gamma(a_0)$ is the total width of the $a_0$.
Then, using the experimental value
$ \Gamma(K_0^*(1430) \rightarrow K \pi ) = 267 \pm 47$
we get $0.21 \pm 0.06 $
for the l.h.s. of Eq.\ (62). A word of caution
is necessary:
the decay $a_0 \rightarrow \eta \pi$ is the dominant
one, but the experimental value of the branching ratio
$a_0 \rightarrow K \bar K$ is still controversial [2].
Given these
uncertainties, the above result, while showing consistency
between theory and experiment, is insufficient for us
to reach any firm conclusion. A better measurement
of the ratio given in Eq.\ (62)
is necessary: a result  clearly
outside the range 0.072-0.144 would rule out a
$q \bar q $ interpretation of the $a_0$.

Next, we consider the decays of the $a_0$ and
$f_0$ into two photons.
For the $a_0$, the experimental result is [2]
$${\Gamma( a_0 \rightarrow \gamma \gamma )
 \Gamma( a_0 \rightarrow
\eta \pi ) \over \Gamma_{\rm tot}( a_0 )  }
= 0.24 \pm 0.08 \ {\rm keV}. \eqno(64)$$
If we use once more the approximation
 $\Gamma(a_0 \rightarrow \eta \pi )
\approx \Gamma( a_0 ) $ we get
$$\Gamma(a_0 \rightarrow \eta \pi )
\approx 0.24 \pm 0.08 \ {\rm keV} . \eqno(65)$$
Similarly, for the $f_0$, one has [2]
$$\Gamma( f_0 \rightarrow \gamma \gamma )  =
0.56 \pm 0.11 \ {\rm keV}. \eqno(66)$$
The results in Eqs.\ (65) and (66)
are not well understood theoretically.
Barnes [88],
assuming that the $a_0$ and $f_0$ are
light-quark mesons and that all the $l=1$ states have
identical spatial wave functions, finds
$\Gamma( a_0 \rightarrow \gamma \gamma )
\approx 1.5$ keV and $\Gamma(f_0 \rightarrow \gamma \gamma)
\approx 4.5$ keV, in disagreement with
the results (65) and (66).
Also a $K \bar K$ state is disfavored because it leads
to [88]
$\Gamma( a_0 \rightarrow \gamma \gamma )=0.6$ keV.
The experimental result (66) appears to be
most compatible with a $q \bar q q \bar q$
hypothesis [89],
for which the suggested  width is  0.27 keV. In
this  case the decay into two photons (of either
the $a_0$ or the $f_0$)
through two colorless vector mesons,  shown in
the  diagram of Fig.\ 2 a),
gives a small result in consequence
of the smallness of their recoupling coefficients as
$ q \bar q q \bar q$ members of the $0^+$ four-quark nonet.
In this scenario, the
decay proceeds mainly through the OZI-suppressed channels,
where one gluon is exchanged (diagrams of Fig.\ 2 b), c)).
Babcock and Rosner [90],
however, find a very different result
$\Gamma( a_0 \rightarrow \gamma \gamma ) \approx 40$ eV
for a quarkonium meson. The theoretical predictions
for quarkonium states are very dispersed [88, 90--95],
as can be seen from Table V.

Interesting information can come from a measurement
of the ratio
$\Gamma( f_0 \rightarrow \gamma \gamma ) /
\Gamma( a_0 \rightarrow \gamma \gamma )$, which,
in the case of $f_0$ and $a_
0$ made of $ q \bar q$ states, can be predicted
quite reliably, as it
depends only on the quark charges and $SU(6)$
Clebsch-Gordan coefficients [88]
 $${\Gamma( f_0 \rightarrow \gamma \gamma ) \over
\Gamma( a_0 \rightarrow \gamma \gamma )} = \biggl
[ {(2/3)^2+(-1/3)^2 \over (2/3)^2-(-1/3)^2} \biggr ]^2
\approx 2.8. \eqno(67)$$
In contrast one expects, in a first
approximation, this ratio to be
$\approx 1$ for multiquark states (whatever their
configuration may be). This follows
because the two  particles decay through the same kind of
processes, namely, according to  the diagrams of
Fig.\ 2 b), c)
in the case of a $q\bar q q \bar q$
state, and according to the diagram of Fig.\ 2 a)
in the case of a
$K \bar K$ molecule, where the decay goes via the
formation of the photons from the
two vector meson components.
The experimental decay ratio $ 2.3 \pm 0.9$
seems to favor the $q\bar q$ interpretation.

The ratio
$\Gamma( f_0 \rightarrow \gamma \gamma ) /
\Gamma( a_0 \rightarrow \gamma \gamma )$
can also be used to get a bound
 on the $s \bar s$ and $G$ components
of the isoscalar particle, assuming that it is
a mixture of $q\bar q$, $s\bar s$ and $G$.
We define $X_S$, $Y_S$ and $Z_S$ for a scalar meson $S$
analogously to our definition for a pseudoscalar
below Eq.\ (3).
For a light-quark meson
mixed with a glueball, we use the experimental value
of the ratio
$${\Gamma( f_0 \rightarrow \gamma \gamma ) \over
\Gamma( a_0 \rightarrow \gamma \gamma )} =
\biggl [ {(2/3)^2+(-1/3)^2
\over (2/3)^2-(-1/3)^2} \biggr ]^2
| X_{f_0} | ^2\approx 2.3
\eqno(68)$$
to find the bound $ 0.71 < | X_{f_0} | < 1$.
If, on the other hand, the decay is dominated by
the $s \bar s$ component we should have
$${\Gamma( f_0 \rightarrow \gamma \gamma ) \over
\Gamma( a_0 \rightarrow \gamma \gamma )} =
\biggl [ {\sqrt{2}  (-1/3)^2
\over (2/3)^2-(-1/3)^2} | Y_{f_0} | \biggr ]^2
\approx 2.3. \eqno(69)$$
This excludes a purely strange meson--glueball
mixing, since it would require
$|Y_{f_0}|$ to be $>2$, violating the theoretical limit
$|Y_{f_0}|<1$.

Pursuing this line of thought further,
we are led to conclude that
the strange component of the $f_0$ cannot be too large;
in general one has:
$${\Gamma( f_0 \rightarrow \gamma \gamma ) \over
\Gamma( a_0 \rightarrow \gamma \gamma )} =
\biggl [ {(2/3)^2+(-1/3)^2
\over (2/3)^2-(-1/3)^2}   | X_{f_0} | \biggr ]^2 +
\biggl [ {\sqrt{2}  (-1/3)^2
\over (2/3)^2-(-1/3)^2} | Y_{f_0} | \biggr ]^2
\approx 2.3.\eqno(70)$$
Note that the $Z_{f_0}$ component is absent because
it is decoupled from the two-photon channel in  lowest
order. We can extract some
information from (70), however, if we use
$X_{f_0}^2+Y_{f_0}^2 \simeq 1$ (i.e. we
neglect the gluonic component).  Then we get
$$ | Y_{f_0} | = 0.4 \pm 0.4 \eqno(71)$$
A more precise measurement of the ratio
$\Gamma(f_0 \rightarrow \gamma \gamma )/
\Gamma( a_0 \rightarrow \gamma \gamma )$
would be very valuable in order
to provide a good test of the models.

Other interesting tests of models in two-photon decays come
from a comparison of the previous widths with that of
the $f_0(1400)$. These tests are not precise because
the $f_0(1400)$ is so much heavier than the $a_0$
and the $f_0$.
Only a rough measurement of the two-photon
decay width of the  $f_0(1400)$ has been made so far [2]:
$ 5.4 \pm 2.3$ keV.
Given that the $f_0(1400)$ is probably an almost
pure $ u \bar u + d \bar d$ state
(it decays mostly into pions)
and taking into account the different kinematical
factors, we expect from Eq.\ (67)
$${\Gamma( f_0(1400) \rightarrow \gamma \gamma ) \over
\Gamma( a_0 \rightarrow \gamma \gamma )} \simeq
2.8 \biggl
({1400 \over 980} \biggr ) ^3   = 8.2, \eqno(72)$$
and, similarly,
$${\Gamma( f_0(1400) \rightarrow \gamma \gamma ) \over
\Gamma( f_0 \rightarrow \gamma \gamma )}
\simeq 3 \eqno(73)$$
if these are  all light-quark resonances.
Experimentally, one finds for these ratios
 $22.5 \pm 12.2$ and $ 9.6 \pm 4.5$ respectively,
which, although not
inconsistent with the predictions of Eqs.\ (72) and (73),
are not in support of them either. Thus, once again
no definitive conclusion can be drawn without better
measurements.  A better determination of
$\Gamma( f_0(1400) \rightarrow \gamma \gamma ) $
would be very useful.
Also the determination of the width of the
$a_0(1320)$ into two
photons  would be interesting. A comparison
of these two partial decay widths
could clarify, through Eq.\ (67), whether these
resonances both belong to the same meson nonet.

We come now to the analysis of a possible
$s \bar s$ component in the $f_0$. With the
assumption of dominance of the spectator quark diagrams [53],
the decay $ B(D_s^+ \rightarrow f_0 \pi^+) =
( 7.8 \pm 3.2 ) 10^{-3}$ indicates the presence of
$s\bar s$.
In the case of an $f_0$ made predominantly by
$s \bar s$, we have
$$ B(D_s^+ \rightarrow f_0 \pi^+) \gg
B(D_s^+ \rightarrow a_0 \pi^+) \eqno(74)$$
and
$$ B(D_s^+ \rightarrow f_0 \pi^+) \gg
B(D^+ \rightarrow f_0 \pi^+), \eqno(75)$$
since the decays in which there is a strange quark
only in the initial
or in the final state are suppressed.
By contrast, in the case of multiquark states,
both the $f_0$ and the $a_0$
have similar composition. Then we expect, in
first approximation,
$$ B(D_s^+ \rightarrow f_0 \pi^+) \approx B(D_s^+
\rightarrow a_0 \pi^+) \eqno(76)$$
and
$$ B(D_s^+ \rightarrow f_0 \pi^+) >
B(D^+ \rightarrow f_0 \pi^+), \eqno(77)$$
where the second inequality is due to the  phase space
difference and to the breaking of $ SU_F (3) $.

In the case that a light-quark component predominates,
we expect:
$$ B(D_s^+ \rightarrow f_0 \pi^+) \ll
B(D^+ \rightarrow a_0 \pi^+
) \approx B(D^+ \rightarrow f_0 \pi^+), \eqno(78)$$
following again the rule that
the decays in which there is a strange quark only
in the initial or in the final state are suppressed.
We also expect
$$ B(D_s^+ \rightarrow f_0 \pi^+) \approx
 B(D_s^+ \rightarrow a_0 \pi^+) \eqno(79)$$
since now both particles are composed of light quarks.
Intermediate situations can, of course,
occur in the case of mixing.
We also note  that a gluonic component should
 couple weakly to $D$ mesons
(one can expect an analogous pattern for $B$ mesons).

The decays of $D^{+}_{s}$, $D^{+}$ and
$D^{0}$ into $f_0 K$
are all suppressed in a quark-antiquark
channel, except for
$D^{0} \rightarrow \bar{K}^{0} f_0$ if
$f_0$ has a large light-quark component.
Recently such a decay has been
analyzed by the ARGUS collaboration [96], who find
$B(D^{0} \rightarrow \bar{K}^{0} f_0) = (0.48 \pm 0.20) \%$.
This branching ratio, while small, is comparable to the
branching ratio of the $D^0$ into
$\bar{K}^{0}$ plus the
$f_0(1400)$, which is believed to be composed
predominantly of light quarks.
(The ARGUS result [96] is
$B(D^{0} \rightarrow \bar{K}^{0} f_0(1400)) =
(0.71 \pm 0.28) \%$.)
Therefore, the observed branching ratio into $f_0$
suggests that its
light-quark component  should not be neglected.

A rough quantitative estimate of the strange component
of the $f_0$ can also be obtained from the
experimental value of the ratio
$$ { B(f_0 \rightarrow \pi \pi ) \over
B( f_0 \rightarrow K \bar K)} \approx 3.56 \eqno(80)$$
in the limit in which $f_0 \rightarrow K \bar K$
proceeds through the strange component  only.
Taking into account threshold effects in the
$K \bar K$ channel by integrating over a Breit--Wigner
multiplied by the appropriate  phase space factor, we get:
$$ \eqalignno{ { B(f_0 \rightarrow \pi \pi )
\over B( f_0 \rightarrow K \bar K)}
= & { (k_{\pi} / m_{f_0})
 \int_0^\infty dE /[ ( 2 (E-E_0) /\Gamma ) ^2 +1]
\over \int_{2 m_K}^\infty dE
\sqrt{ (E^2-4 m_K^2)/ (4 E^2)} /
[ 2
(E-E_0) / \Gamma )^2 +1 ]}
 \biggl \vert { X_{f_0} \over Y_{f_0}} \biggr \vert ^2
\cr
\approx  & 10 \biggl \vert { X_{f_0} \over Y_{f_0}}
 \biggr \vert ^2. & (81) \cr} $$
If there is  no gluonic contribution to the decay ratio
(80), then
$$ X_{f_0}^2+Y_{f_0}^2=1.\eqno(82)$$
If we substitute Eq.\ (82) into Eq.\ (81) and use Eq.\ (80),
we obtain:
$$| Y_{f_0} | = 0.86. \eqno(83)$$
Of course, the light-quark component can also
decay into $ K \bar K$,
and therefore the above value of $| Y_{f_0} |$
has to be considered as an upper bound.
A lower bound
on $| Y_{f_0} |$
can be obtained in the limit of
unbroken $SU_F (3)$; in such a case the
$ u \bar u + d \bar d $ component will decay
with  2/3 probability into
$ \pi \pi$ and 1/3 into $K \bar K$.
We get, therefore, from Eq.\ (81):
$$ { B(f_0 \rightarrow \pi \pi )
\over B( f_0 \rightarrow K \bar K)}
= 10  { 2/3 | X_{f_0} |^2 \over | Y_{f_0} |^2+
1/3 | X_{f_0} |^2} \approx 3.56, \eqno(84)$$
which leads to $| Y_{f_0} |=0.78$.

If the $f_0$ were a pure gluonic state,
its decays should be in accord with flavor democracy.
This, with the usual assumption of an unbroken
$SU_F (3)$ hypothesis, leads to a much too large result:
$$ { B(f_0 \rightarrow \pi \pi ) \over
B( f_0 \rightarrow K \bar K)}
\approx 10 \times { 4 \over 5} =  8. \eqno(85)$$
The experimental value given in Eq.\ (80) suggests
that the $f_0$ does not have a large gluonic component.

As we have just seen,
the above argument  (although obtained with
rough approximations) leads to the result that
$0.78 \lsim Y_{f_0} \lsim 0.86$, not in good agreement
with the smaller value of $Y_{f_0}$
obtained earlier [see Eq.\ (71)].
Granting that in both cases the
arguments are {\it very} qualitative,
it is clear that better experimental
measurements will be necessary to clarify our picture
of what is going on.
Of course, a better theoretical
understanding of the general situation is also desirable.

Another test of the strange component in the $f_0$
could come from the high energy behavior of
$\pi^- p \rightarrow f_0 n$. We expect
$${\sigma(\pi^- p \rightarrow f_0 n ) \over
\sigma(\pi^- p \rightarrow a_0 n ) } \approx 1 \eqno(86)$$
in the case that
$a_0$ and $f_0$ are both light-quark mesons, because both
processes proceed through the exchange of the same
Regge trajectory (either
the $\pi$ or the $a_1$ trajectory). Also for a multiquark
system one can expect a ratio near one.
In contrast, the ratio (86) could be OZI suppressed if
the $s \bar s$ component of the $f_0$ is important.
Then the ratio (86) could be
lowered by a factor $\sim \alpha_s ^2 \approx 0.06$.
If the gluonic component dominates, the ratio (86) could be
 $\sqrt{OZI}$ suppressed and  could be
lowered by a factor  $\sim  \alpha_s \approx 0.2$.
(See, however, Chanowitz [97] for a discussion of the
 $\sqrt{OZI}$ suppression).
Yet another test of the $s \bar s$ component of the
$f_0$ comes
from the comparison of $\sigma(\pi^- p \rightarrow f_0 n ) $
and $\sigma(K^- p \rightarrow f_0 \Lambda [\Sigma^0] ) $.
If the $f_0$ is mainly $s \bar s$ then the first channel is
proportional to $\alpha_s^2$ and the second to $\alpha_s$.
It follows that at sufficiently high energy, we have
$${\sigma(\pi^- p \rightarrow f_0 n ) \over
\sigma(K^- p \rightarrow f_0 \Lambda ) } < 1~.\eqno(87) $$
In contrast, if the $f_0$ is mainly a light-quark
state,  we expect:
$${\sigma(\pi^- p \rightarrow f_0 n ) \over
\sigma(K^- p \rightarrow f_0 \Lambda ) } > 1,\eqno(88) $$
since the first process proceeds through the exchange of
the $\pi$ Regge trajectory ($\alpha_\pi(0) \approx 0$)
and the second through the exchange of the K trajectory
($\alpha_K (0) \approx -0.3$).

We next give some considerations on a possible gluon
component of the $f_0$.
The observed branching ratio of the $J/\psi$
into $\omega f_0$ is
 $( 1.4 \pm 0.5 )\times 10^{-4}$ and
 $( 3.2 \pm 0.9 )\times 10^{-4}$ into $\phi f_0$
(into $\rho a_0 $ has been not observed yet).
Both disagree with the hypothesis of the $f_0$
being predominantly $s \bar s$, which has as consequences:
$$B(J/\psi \rightarrow \omega f_0) \approx 0, \eqno(89)$$
being a doubly OZI violating decay, and
$$B(J/\psi \rightarrow \rho^0 a_0^0) \approx
B(J/\psi \rightarrow \phi f_0)\eqno(90)$$
(neglecting phase space and $SU_F (3)$ breaking effects)
being both singly OZI forbidden decays.

The lack of the observation of the decay
$J/\psi \rightarrow \rho a_0$
is also in disagreement with the molecular
hypothesis according to which [98]
$$B(J/\psi \rightarrow \rho^0 a_0^0) =
{1 \over 2} B(J/\psi \rightarrow \phi f_0)
=B(J/\psi \rightarrow \omega f_0)\eqno(91)$$
The $K \bar K$ molecule predictions should be
relatively unaffected by the the $SU_F (3)$
breaking. By contrast, in the $ q \bar q$
hypothesis $SU_F (3)$ breaking effects are expected
to suppress
$J/\psi \rightarrow \phi f_0$, analogously to
what is observed for
the tensor mesons, where [99]
$ B(J/\psi \rightarrow \rho^0 a_2^0) \approx
5  B( J/\psi \rightarrow \phi f_2')$.

Of course, different mixing schemes can produce
quite different results, but if the decays
of the $f_0$ proceed through a gluonic component
this would explain the lack of observation of the
$a_0$; the latter
 having no gluonic component, a much smaller
$B(J/\psi \rightarrow \rho^0 a_0^0)$ is expected.
An unusually large gluonic component would also lead to
(neglecting phase space differences)
$$B(J/\psi \rightarrow \omega f_0) \approx
B(J/\psi \rightarrow \phi f_0). \eqno(92)$$
if the $f_0$ is formed through its gluonic part.

A bound on the gluonic component of the
$f_0$ comes from comparing
$$B(J/\psi \rightarrow \gamma f_0) \ B( f_0
 \rightarrow \pi \pi ) < 7 \times 10^{-5} \eqno(93)$$
with
$$B(J/\psi \rightarrow \gamma f_0(1710)) \ B( f_0(1710)
 \rightarrow K \bar K ) = ( 9.7 \pm 1.2 )\times 10^{-4}.
\eqno(94)$$
(However, the large mass
difference between the two final states should be matter of
caution.) If the decay proceeds predominantly  through a
gluonic component, we  have:
$${B(J/\psi \rightarrow \gamma f_0) \over
B(J/\psi \rightarrow \gamma f_0(1710))} =
\biggl \vert {Z_{f_0}
\over Z_{f_0(1710)}} \biggr \vert ^2 \biggl
\vert { k_{f_0}
\over k_{f_0(1710)}} \biggr \vert ^3 = 2.18  \biggl
\vert {Z_{f_0} \over Z_{f_0(1710)}} \biggr \vert ^2,
\eqno(95)$$
where $| k_{f_0} / k_{f_0(1710)}|^3$
comes from  phase space. The above
implies $$\biggl \vert {Z_{f_0}
\over Z_{f_0(1710)}} \biggr \vert ^2 < 4 \times 10^{-2}.
\eqno(96)$$
For a  purely gluonic $f_0(1710)$ we then have
$\vert Z_{f_0} \vert < 0.2.$
The data, however, are still quite doubtful;
a precise determination
of these branching ratios and  of those of the $f_0(1590)$
and of the $f_0(1400)$ would give information on their
gluonic component.
However, even if such a component is negligible for one
 or all of these particles,
these branching ratios would give a clue on their
quarkonic composition. For instance, in the case $Z_A
\approx Z_B \simeq 0$, we have
$${B(J/\psi \rightarrow \gamma A) \over
B(J/\psi \rightarrow \gamma B)} =  \biggl \vert { k_A
\over k_B} \biggr \vert ^3  \biggl
\vert {\sqrt{2} X_A + Y_A \over
\sqrt{2} X_B + Y_B} \biggr \vert ^2. \eqno(97)$$
Then  we find
$${B(J/\psi \rightarrow \gamma f_0) \over
B(J/\psi \rightarrow \gamma f_0(1400))} = 1.45 \eqno(98)$$
if both particles are composed of light quarks. On the
other hand, this ratio is $=0.73$ if
the more massive state is composed predominantly
of light quarks and the lighter one  has a predominantly
strange-quark component, as some of the previous
considerations seem to indicate.
See {\it e.g.} Eq.\ (83) and also the discussion preceding
Eq.\ (74)).

\bigskip

\centerline{IV. CONCLUSIONS}
\bigskip

Despite a vast literature on the subject of spin-0 mesons,
the subject is not really well
understood either theoretically or experimentally.
We would like to know the quark and glueball content
of these mesons
(and their four-quark and hybrid content as well), but
we believe that it will take
the next generation of experiments,
together with additional theoretical work, before we
have a reasonably accurate picture.
It is for this reason
that we have tried to gather together in one place not
only information
about the present status  of the subject, but also a
large number of suggestions
about feasible measurements which should help to clarify
the situation.

Concerning the pseudoscalar sector, we have presented a
simple mixing scheme (involving the $\eta$,
the $\eta'$ and the $\iota$) and proposed various
ways to test the composition of these particles.
We believe that no single measurement will solve
the complicated puzzles offered by these mesons, and we
have suggested a number of different measurements which
should help us to arrive at a
definite solution. Such measurements should be well
within experimental reach during the next few years.

The present picture is even less clear for the
scalar mesons, where many poorly known
resonances are involved.
Nevertheless, also in this case it appears that a number of
measurements we have suggested could help
considerably in clarifying the situation.

\bigskip
\centerline{ACKNOWLEDGMENTS}
We thank Alex Dzierba, Ben Brabson, and Stuart Samuel
for helpful discussions. This work
was supported in part by the Italian Ministry of
University Scientific Research and Technology (MURST)
and in part by the U.S. Department of Energy.
\bigskip

\vfill\eject

\centerline{\bf References}
\medskip

\item{[1]} C. A. Heusch, Nucl.\ Phys.\ B (Proc. Suppl.)
{\bf 27}, 263 (1992);
M. R. Pennington, in {\it Plots, Quarks
and Strange Particle}, edited by I. J. R. Aitchison, C. H.
Llewellyn Smith, and J. E. Paton, World Scientific
(1991), p. 66;
M. Burchell, in {\it XII Intern.\ Conf.\ on the
Physics of Collisions}, Boulder, CO, 1992 (proc. to be
published).

\item{[2]} Particle Data Group, K. Hikasa {\it et al.},
Phys.\ Rev.\ D {\bf 45}, S1 (1992).

\item{[3]} E. V. Shuryak, Nuclear Phys.\ {\bf B214}, 237
(1983).

\item{[4]} W. H. Blask {\it et al.}, Z. Phys.\ A {\bf 337}, 327
(1990).

\item{[5]} A. I. Vainshtein, V. I. Zakharov,
V. A.  Novikov, and M. A.  Shifman,  Sov.\ J. Part.\ Nucl.\
{\bf 13}, 224 (1982).

\item{[6]}
E. V. Shuryak, {\it The QCD vacuum, hadrons, and 
superdense matter} (World Scientific, Singapore, 1988).

\item{[7]} F. Caruso, E. Predazzi, A. C. B. Antunes and
J. Tiomno,  Z.\ Phys.\ C {\bf  30}, 493 (1986)
and references therein.

\item{[8]} Y. A. Simonov, Nucl.\ Phys.\  B (Proc. Suppl.)
{\bf 23}, 283 (1991).

\item{[9]} T. Teper, {\it Intern.\ European Conf. on High
Energy Phys., Brighton } (1983);
K. Ishikawa, M. Teper, and G. Schierholz, Phys.\ Lett.\ B
{\bf 116}, 429 (1982).

\item{[10]} C. Michael and M. Teper, Phys.\ Lett.\ B
{\bf 206}, 299 (1988).

\item{[11]} C. Michael and M. Teper, Nucl.\ Phys.\
{\bf B314}, 347 (1989).

\item{[12]} V. A. Novikov {\it et al.}, Acta\ Phys.\ Pol.\
 {\bf B12}, 399 (1981).


\item{[13]} M. Chanowitz and S. Sharpe,  Nucl.\ Phys.\
{\bf B222}, 211 (1983).

\item{[14]} K. Ishikawa {\it et al.}, Z. Phys.\ C {\bf 21},
167 (1983)).

\item{[15]}
J. Cornwall and A. Soni, Phys.\ Lett.\ B {\bf 120},
431 (1983).

\item{[16]} B. Berg and A. Billoire, Nucl.\ Phys.\
{\bf B221}, 109 (1983).

\item{[17]} S. Narison, Z.\ Phys.\ C\ {\bf 26},
204 (1984).

\item{[18]} N. Isgur and J. Paton, Phys.\ Rev.\ D
{\bf 31}, 2910 (1985).

\item{[19]} K. Geiger, B. M\"uller, W. Greiner, Z. Phys.\ C
{\bf 48}, 257 (1990).

\item{[20]} S. Iwao, Lett. Nuovo Cimento {\bf 32},
475 (1981); S. Iwao, {\it Quarks 82, Sukhumi Proc.}
Moscow Editions (1983), p.\ 228.

\item{[21]} A. S. de Castro and H. F. de Carvalho,
J. Phys.\ G {\bf 16}, L81 (1990).

\item{[22]} S. Bhatnagar and A. N. Mitra,
 Nuovo Cimento A {\bf 104}, 925 (1991).

\item{[23]} M. Anselmino, M. Genovese and E. Predazzi,
Phys. Rev. D {\bf 44}, 1597 (1991).

\item{[24]} S. Godfrey and N. Isgur,
Phys.\ Rev.\ D {\bf 32}, 189 (1985).

\item{[25]} H. Goldberg, Phys.\ Rev.\ Lett.\
{\bf 44}, 363 (1980).

\item{[26]} J. Jousset {\it et al.},\ Phys. Rev. D {\bf 41},
1389 (1990).

\item{[27]} J. Perrier in {\it IV Conf. on "Physics in
Collision"}, Santa Cruz, edited by A. Seiden
(Editions Frontieres, France, 1984), p. 143.

\item{[28]} S. S. Pinsky,  Phys.\ Rev.\ D {\bf 31 }, 1753
(1985).

\item{[29]}D. Coffman {\it et al.}, Phys.\ Rev.\ D {\bf 38},
2695 (1988).

\item{[30]} N. Aizawa {\it et al.},  Progr.\ Theor.\ Phys.
(Kyoto) {\bf 62}, 1085 (1979).

\item{[31]} C. Rosenzweig, A. Salomone, and J. Schechter,
Phys.\ Rev.\ D {\bf 24}, 2545, 1981.

\item{[32]} M. Frank and P. J. O'Donnell, Phys.\ Lett.\ B
{\bf 144}, 451 (1984);
Phys.\ Rev.\ D {\bf 32}, 1739 (1985).


\item{[33]} S. Basu, and B. Bagchi, Z. Phys.\ C
{\bf 37}, 69 (1987); Mod.\ Phys.\ A {\bf 6}, 633 (1988);
S. Basu, A. Lahiri, and B. Bagchi,
Phys.\ Rev.\ D {\bf 37}, 1250 (1988).

\item{[34]} H. Genz, M. Nowakowski, and D. Woitschitzky,
Phys.\ Lett.\ B {\bf 250}, 143 (1990).


\item{[35]} N. Morisita, I. Kitamura, and T. Teshima,
Phys.\  Rev.\ D {\bf 44}, 175 (1991).

\item{[36]} J. Bartelski and S. Tatur,
Phys.\ Lett.\ B {\bf 289}, 429 (1992) and
references therein.

\item{[37]} M. G. Rath {\it et al.},
Phys.\ Rev.\ D {\bf 40}, 693 (1989).

\item{[38]} S. Blessing {\it et al.}, in {\it Glueballs,
Hybrids
and Exotic Hadrons}, edited by S.-U. Chung, AIP Conf.\
Proc.\ No.\ 185 (Particles and Fields Series No.\ 36),
(1987) p.\ 363.

\item{[39]} G. Marchesini and B. R. Webber, Nucl.\ Phys.\
{\bf B310}, 461 (1988);
I. G. Knowels,  Nucl.\ Phys.\ {\bf B310}, 571 (1988);
G. Marchesini {\it et al.}, Comp.\ Phys.\ Comm.\
{\bf 67}, 465 (1992).

\item{[40]} T. Sj\"ostrand, Comp.\ Phys.\ Comm.\
{\bf 39}, 347 (1986); T. Sj\"ostrand and M. Bengtsson,
Comp.\ Phys.\ Comm.\ {\bf 43}, 367 (1987).

\item{[41]} P. Chliapnikov, private communication.

\item{[42]} H. Albrecht {\it et al.}, DESY 92-174 (1992).

\item{[43]} B. Andersson {\it et al.}, Phys.\ Rep.\
{\bf 97}, 31 (1983).

\item{[44]} C. D. Buchanan and S. B. Chun, Phys.\ Rev.\
Lett.\ {\bf 59}, 1997 (1987).

\item{[45]} S. Klimt {\it et al.},
Nucl.\ Phys.\ {\bf A516}, 429 (1990).

\item{[46]} J. F. Donoghue, B. R. Holstein and Y.-C.\ R.
Lin, Phys.\ Rev.\ Lett.\ {\bf 55}, 2766 (1985);
J. Gasser and H. Leutwyler, Nucl.\ Phys.\ {\bf B250},
465 (1985).

\item{[47]} R. Sommer, Nucl.\ Phys.\ B
(Proc.\ Suppl.) {\bf 17}, 513 (1990).

\item{[48]} H. E. Haber and J. Perrier,
Phys.\ Rev.\ D {\bf 32}, 2961 (1985).

\item{[49]} C. Amsler {\it et al.},
Phys.\ Lett.\ B {\bf 294}, 451 (1992).

\item{[50]}  M. S. Chanowitz, Phys.\ Rev.\ Lett.
{\bf 35}, 977 (1975);  Phys.\ Rev.\ Lett.\ {\bf 44},
59 (1980).

\item{[51]} M. Feindt,  CERN--PPE/92--148 (1992);
M. Benayoun {\it et al.},  CERN--PPE/92--156 (1992).

\item{[52]} J. L. Rosner, Phys.\ Rev.\ D {\bf 27},
1101 (1983).

\item{[53]}   I. Bediaga, F. Caruso, and E. Predazzi,
Nuovo Cimento A {\bf 91}, 306 (1986).

\item{[54]} E. Leader and  E. Predazzi,
{\it An Introduction to Gauge Theories
and the New Physics}, Cambridge Univ.\ Press (1983).

\item{[55]} V. Elias and M. D. Scadron, Phys.\ Rev.\ D
{\bf 30}, 647 (1984); D. I. Diakonov and  V. Y. Petrov,
Nucl.\ Phys.\  {\bf B27}, 457 (1986);
H. D. Politzer, Phys.\ Lett.\ B {\bf 116}, 171 (1982);
Nucl.\ Phys.\ {\bf B117}, 397 (1976).

\item{[56]} S. Rudaz, Phys.\ Rev.\ D {\bf 14}, 298 (1976);
H. Kowalsky and T. F.  Walsh, {\it ibid.} {\bf 14}, 852
(1976) 852; J.  Pasupathy and C. A. Singh,
{\it ibid.} {\bf 18}, 791 (1978).

\item{[57]} F.~E.~Low, Phys.\ Rev.\ {\bf 120}, 582 (1960).

\item{[58]} P. Jenni {\it et al.}, Phys.\ Rev.\ D\ {\bf 27},
1031 (1983).

\item{[59]} R. L. Jaffe, Phys.\ Rev.\ D {\bf 25}, 267
(1977).

\item{[60]} B. Diekmann, Phys.\ Rep.\ {\bf 159}, 99 (1988)
and references therein.

\item{[61]} J. Weinstein and  N. Isgur,
Phys.\ Rev.\ Lett.\ {\bf 48}, 659 (1982);
Phys.\ Rev.\  D {\bf 27}, 588 (1983).

\item{[62]} N. A. T\"ornqvist, Phys.\ Rev.\ Lett.\ {\bf 49},
624 (1982).

\item{[63]} E. Halyu, Phys.\ Lett.\ B {\bf 271}, 415 (1991).

\item{[64]} V. N. Gribov, LU--TP 91--7 (1991).

\item{[65]} F. Binon {\it et al.}, Nuovo\ Cimento\ A
{\bf 78},
313 (1983); Nuovo\ Cimento\ A {\bf 80}, 363 (1984);
D. Alde {\it et al.}, Nucl.\ Phys.\ {\bf B269},
485 (1986); Phys.\ Lett.\ B {\bf 198}, 286 (1987);
Phys.\ Lett.\ B {\bf 201}, 160 (1987);
Y. D. Prokoshkin, Sov.\ Phys.\ Doklady\ {\bf 36}, 155 (1991).

\item{[66]} J. F. Donoghue, in {\it Proc.\ of the Intern.\
Conf.\ Europhysics on High-Energy
Phys.\ Bari, Italy} (1985), p. 326; S. Cooper,
{\it ibid.}, p. 974 (1985).

\item{[67]} M. Genovese, Phys.\ Rev.\ D {\bf 46},
5204 (1992).

\item{[68]} D. Alde {\it et al.}, Nucl.\ Phys.\ {\bf B269},
485 (1986).

\item{[69]} J. Lanik, Z.\ Phys.\ C {\bf 39}, 143 (1988).

\item{[70]} P. van Baal and A. Kronfeld,  Nucl.\ Phys.\  B
(Proc.\ Suppl.) {\bf 9 }, 227 (1990).

\item{[71]} F. Gliozzi, Acta\ Phys.\ Polonica B
{\bf 23}, 971 (1992).

\item{[72]} E. E. Boos and A. U. Daineka,
Phys.\ Lett.\ B {\bf 283}, 113 (1992).

\item{[73]} M. Albanese, Phys.\ Lett.\ B {\bf 192},
163 (1987).

\item{[74]} K. M. Bitar {\it et al.}, Phys.\ Rev.\ D
{\bf 44}, 2091 (1991); ANL-HEP-CP-92-113 (1992).

\item{[75]}
M. Svec {\it et al.}, Phys.\ Rev.\  D {\bf 46}, 948 (1992).

\item{[76]} F. E. Close, N. Isgur, and S. Kumano,
RAL-92-026 (1992).

\item{[77]} R. Kokoski and N. Isgur, Phys.\ Rev.\ D
{\bf 35}, 907 (1987).

\item{[78]} J. Weinstein and N. Isgur,  Phys.\ Rev.\ D
{\bf 41}, 2236 (1990).

\item{[79]} E. S. Swanson,
Ann.\ Phys.\ {\bf 220}, 73 (1992).

\item{[80]} J. Speth {\it et al.}, in {\it XII Particles
and Nuclei Intern.\  Conference, MIT}, (1990).

\item{[81]} D. Morgan and  M. R. Pennington,
Phys.\ Lett.\ B {\bf 258}, 444 (1991);
RAL-92-070 (1992).

\item{[82]} S. Chakrabarty {\it et al.},  Progr.\ Part.\
Nucl.\ Phys.\ {\bf 22}, 43 (1989) and references therein.

\item{[83]} P. Abreu {\it et al.}, Phys.\ Lett.\ B {\bf 298},
236 (1993).

\item{[84]} S. Abachi {\it et al.}, Phys.\ Rev.\ Lett.\
{\bf 57}, 1990 (1986);
Phys.\ Lett.\ B {\bf 199}, 151 (1987).

\item{[85]}  M. Aguillar-Benitez {\it et al.},
Z. Phys.\ C {\bf 50 }, 405 (1991).

\item{[86]} J. Apsimon {\it et al.}, Z.\ Phys.\ C {\bf 56},
185 (1992).

\item{[87]} T. Teshina and  S. Oneda, Phys.\  Rev.\ D
{\bf 33}, 1974 (1986).

\item{[88]}
T. Barnes,  Phys.\ Lett.\ B {\bf 165}, 434 (1985).

\item{[89]} N. N. Achasov {\it et al.}, Phys.\ Lett.\ B
{\bf 108}, 134 (1982).

\item{[90]} J. Babcock and J. L. Rosner, Phys.\ Rev.\ D
{\bf 14}, 1286 (1976).

\item{[91]} S. B. Berger and B. T. Feld,
Phys.\ Rev.\ D {\bf 8}, 3875 (1973).

\item{[92]} A. Bramon and M. Greco,
Lett.\ Nuovo Cimento {\bf 2}, 522 (1971).

\item{[93]} V. M. Budnev and A. E. Kaloshin,
Phys.\ Lett.\ B {\bf 86}, 351 (1979).

\item{[94]} S. Eliezer,  J.\ Phys.\ G {\bf 1}, 701 (1975).

\item{[95]} G. K. Greenhut and G. W. Intemann,
Phys.\ Rev.\ D {\bf 18}, 231 (1978).

\item{[96]} H. Albrecht {\it et al.}, DESY 93-003 (1993).

\item{[97]}
M.S. Chanowitz, in {\it Proc.\ 1981 SLAC Summer Institute on
Particle Physics}, edited by
A. Mosher, SLAC REP 245 p.\ 41 (1982).

\item{[98]} A. Seiden, in {\it Proc.\ VII
Intern.\ Workshop on Photon-Photon Collis\-ions},
(Par\-is, 1986), edited by A. Coureau and P. Kessler,
World Scientific (1986), p.\ 193.

\item{[99]} L. K\"opke and  N. Wermes,
Phys.\ Rep.\ {\bf 174}, 67 (1989).

\vfill\eject
\item{Table I}. Mass predictions for the lowest
pseudoscalar glueball states. These are subdivided into
lattice and flux tube results and results in the
digluonium and trigluonium approximations.

$$\vbox {\halign {\hfil #\hfil &&\quad \hfil #\hfil \cr
\cr \noalign{\hrule}
\cr \noalign{\hrule}
\cr
&Reference \hfil&&Mass (MeV) &\cr
\cr \noalign{\hrule}
\cr
& ({\bf Lattice, Flux Tube}) &&&\cr
&Teper 82 [9] &&1450&\cr
&Berg 83 [16]&&2175&\cr
&Ishikawa 83 [14] &&1250 - 1660&\cr
&Isgur 85 [18] &&2790&\cr
&Michael 88 [10] &&2400&\cr
&Michael 89 [11]&&2464-3124&\cr
&Simonov 91 [8]&&1640&\cr
& ({\bf Digluonium})&& &\cr
&Novikov 81 [12] &&2000 - 2500&\cr
&Chanowitz 83 [13]&&1240 - 1440&\cr
&Cornwall 83 [15] &&1300 - 1400&\cr
&Narison 84 [17] &&$< 1900 \pm 400$ &\cr
&Geiger 90 [19]&&1440&\cr
&({\bf Trigluonium})&& &\cr
&Iwao 83, Iwao 81 [20] && 3310,3620 &\cr
&de Castro 90 [21] && 3200 (1S)&\cr
&Bhatnagar 91 [22] && $\gsim 1541$&\cr
&Simonov 91 [8]&&1800&\cr
\cr \noalign{\hrule}
\cr \noalign{\hrule}
}}$$


\item{Table II}. Scalar resonances from the Review of
Particle Properties [2]. A question mark indicates a
resonance which is not well established.

\vskip 12pt
$$\vbox {\halign {\hfil #\hfil &&\quad \hfil #\hfil \cr
\cr \noalign{\hrule}
\cr \noalign{\hrule}
\cr
&Name of the resonance&&mass&&$\Gamma$ &\cr
& && (MeV)&& (MeV)&\cr
\cr \noalign{\hrule}
\cr
&$f_0(975)$&&$974.1 \pm 2.5$&& $47 \pm 9$&\cr
&$a_0(980)$&&$982.7 \pm 2.0 $&& $57 \pm 11$&\cr
&$f_0(1240)$ ?&&$1240 \pm 30 $&& $140 \pm 30$&\cr
&$a_0(1320)$ ?&&$1322 \pm 30 $&& $130 \pm 30$&\cr
&$f_0(1400)$&&$\approx 1400$&& $150-400$&\cr
&$K^*_0(1430)$&&$1429 \pm 9 $&& $287 \pm 31$&\cr
&$f_0(1525)$ ?&&$\approx 1525$ && $\approx 90$ &\cr
&$f_0(1590)$ ? && $1587 \pm 11$ &&     $175 \pm 19$ &\cr
&$f_0(1710)$ ?&&$1709 \pm 5 $&& $146 \pm 12$&\cr
&$X(1740)$ ?&&$1744 \pm 15 $&& $< 80$&\cr
&$K^*_0(1950)$ ?&&$1945 \pm 30 $&& $201 \pm 103$&\cr
\cr \noalign{\hrule}
\cr \noalign{\hrule}
}}$$

\vfill \eject

\item{Table III}. Mass predictions for the lowest scalar
states. These are subdivided into lattice, flux tube and
string results and results in the digluonium approximation.
\vskip 12pt
$$\vbox {\halign {\hfil #\hfil &&\quad \hfil #\hfil \cr
\cr \noalign{\hrule}
\cr \noalign{\hrule}
\cr
&Reference &&Mass (MeV) &\cr
\cr \noalign{\hrule}
\cr
& ({\bf Lattice, Flux tube, String})&& &\cr
&Teper 82 [9] &&770&\cr
&Berg 83 [16] &&750&\cr
&Ishikawa 83 [14] &&$740 \pm 90$&\cr
&Isgur 85 [18] &&1520&\cr
&Albanese 87 [73] &&650-820&\cr
&Michael 89 [11] &&1408-1672&\cr
&van Baal 89 [70] &&$1370 \pm 90$&\cr
&Halyu 91 [63]&&$\approx 880$ &\cr
&Bitar 91 [74]&&$1200 \pm 300$&\cr
&Gliozzi 92 [71]&& 1542 &\cr
& ({\bf Digluonium})&& &\cr
&Chanowitz 83 [13] &&670-1560&\cr
&Cornwall 83 [15] &&1100-1200&\cr
&Narison 84 [17] &&1400&\cr
&Lanik 88 [69] && 850-990&\cr
&de Castro 90 [21] && 993 (1S)&\cr
&Geiger 90 [19] &&796-1082&\cr
&Boos 92 [72] &&1400&\cr
\cr \noalign{\hrule}
\cr \noalign{\hrule}
}}$$


\item{Table IV}. Predictions of Close {\it et al.}\ [76]
for the absolute branching ratios of $\phi$
into $a_0(980) \gamma$ or into $f_0(975) \gamma$, and
for the ratio of these for different compositions
of the two scalar states.

\vskip 12pt
$$\vbox {\halign {\hfil #\hfil &&\quad \hfil #\hfil \cr
\cr \noalign{\hrule}
\cr \noalign{\hrule}
\cr
&Scalar meson constitution&&Absolute branching ratio
&&$\Gamma(\phi \rightarrow a_0 \gamma) \over
\Gamma(\phi \rightarrow f_0 \gamma)$&\cr
\cr \noalign{\hrule}
\cr
&$K \bar K$ molecule&& $a_0 \approx f_0 \approx 4 \times
10^{-5}$&&$\approx 1$&\cr
&$q \bar q q \bar q~~  K \bar K$ bag&&  $< 10^{-6}$&&
$ 1$&\cr
&$q \bar q q \bar q~~  D \bar D $ bag&& $< 10^{-6}$
&&9&\cr
&$q \bar q q \bar q~~  (u \bar u + d \bar d)
(s \bar s)$ bag&& $< 10^{-6}$&&--&\cr
&$f_0$ glueball, $a_0$ quarkonium&&  $\lsim 10^{-6}$&&
$\approx 1$&\cr
&$f_0$ and $a_0$ light quark $^3 P_0$&&  $\lsim 10^{-6}$
&&$\approx 1$&\cr
&$f_0 ~s \bar s$ and $a_0$ light quark $^3 P_0$
&& $f_0 \lsim 10^{-5}$ &&$\approx 0$&\cr
\cr \noalign{\hrule}
\cr \noalign{\hrule}
}}$$

\vfill \eject

\item{Table V}. Predictions of the width into
two photons for the $a_0(980
)$ and the $f_0(975)$ in different models.

\vskip 12pt
$$\vbox {\halign {\hfil #\hfil &&\quad \hfil #\hfil \cr
\cr \noalign{\hrule}
\cr \noalign{\hrule}
\cr
Reference && Model
&& $\Gamma(a_0 \rightarrow \gamma \gamma) $
&& $\Gamma(f_0 \rightarrow \gamma \gamma) $&\cr
 && && (keV) && (keV)&\cr
\cr \noalign{\hrule}
\cr
Babcock 76 [90] &&Vector dominance && 0.04&&
8&\cr
Barnes 85 [88] && $f_0$ Light quarks && 1.5&&
4.5&\cr
Berger 73 [91] &&Vector dominance && 3.8&&
--&\cr
Bramon 71 [92] && Reggeon exchange && 50&&
--&\cr
Budnev  79 [93] && Potential && 4.8&&
$12.8B$*&\cr
 &&(Harmonic oscillator)&& && &\cr
Eliezer 75 [94] && $a_0 \rightarrow \gamma \gamma $ through
&& --&&
0.2&\cr
 && (2 pseudoscalars) && && &\cr
Greenhut 78 [95]&&Vector dominance && $550 \pm 270$&&
--&\cr
\cr \noalign{\hrule}
\cr \noalign{\hrule}
}}$$

*The quantity $B$ depends on the quark content of the $f_0$.

\vskip 3cm
\noindent{\bf Figure captions}
\medskip

Fig. 1.  The process $e^+ e^- \rightarrow e^+ e^- +
{\rm pseudoscalar}$
in the equivalent photon approximation.
The symbol $P$ is for pseudoscalar meson.

Fig. 2. Possible mechanisms of decay of the $a_0(980)$
and $f_0(975)$ mesons in models in which these mesons
contain two quarks and two antiquarks.
The symbol $V$ is for vector meson.

\bye